\documentclass[12pt,preprint]{aastex}

\def\la{\mathrel{\mathpalette\fun <}}
\def\ga{\mathrel{\mathpalette\fun >}}
\def\simpropto{\lower.2ex\hbox{$\; \buildrel \sim    \over \propto \;$}}

\def\fun#1#2{\lower0.837ex\vbox{\baselineskip0ex\lineskip0.209ex
  \ialign{$\mathsurround=0ex#1\hfil##\hfil$\crcr#2\crcr\sim\crcr}}}

\def\msun{M_\odot}
\def\msunyr{M_\odot \ {\rm yr}^{-1}}

\def\sles{\lower2pt\hbox{$\buildrel {\scriptstyle <}
   \over {\scriptstyle\sim}$}}

\def\sgreat{\lower2pt\hbox{$\buildrel {\scriptstyle >}
   \over {\scriptstyle\sim}$}}

\def\la{\mathrel{\mathpalette\fun <}}
\def\ga{\mathrel{\mathpalette\fun >}}

\begin{document}

\title{The {\it Kepler} Light Curve of V344 Lyrae: 
   Constraining the Thermal-Viscous Limit Cycle Instability} 
  
\shortauthors{CANNIZZO ET AL}
\author{ 
         J.~K.~Cannizzo\altaffilmark{1,2},
         M.~D.~Still\altaffilmark{3,4},
         S.~B.~Howell\altaffilmark{5},
         M.~A.~Wood\altaffilmark{6},
          and
         A.~P.~Smale\altaffilmark{2}
}
\altaffiltext{1}{CRESST/Joint Center for Astrophysics,
                 University of Maryland, Baltimore County, Baltimore, MD 21250,
                 John.K.Cannizzo@nasa.gov}
\altaffiltext{2}{Astroparticle Physics Laboratory, 
                 NASA-Goddard Space Flight Center, Greenbelt, MD 20771}
\altaffiltext{3}{Bay Area Environmental Research Inst., Inc.,
                 560 Third St. W, Sonoma, CA 95476,
                 Martin.D.Still@nasa.gov}
\altaffiltext{4}{NASA-Ames Research Center, Moffet Field, CA 94035}
\altaffiltext{5}{National Optical Astronomy Observatory, Tucson, AZ 85719}
\altaffiltext{6}{Department of Physics and Space Sciences, Florida Institute of 
              Technology, 150 W. University Blvd., Melbourne, FL 32901}
\begin{abstract}
   We present time
dependent modeling based on the accretion disk 
 limit cycle model  for  a 270 d light curve of
  the short period SU UMa-type
   dwarf nova   V344 Lyr taken by {\it Kepler}.
  The unprecedented precision  
 and  cadence (1 minute)
   far surpass that generally available
for long term light curves.
 The data encompass two superoutbursts and 17 normal
  (i.e., short)
outbursts.
  The  main decay of the superoutbursts  is
   nearly perfectly
   exponential,
  decaying
  at a rate $\sim$12 d mag$^{-1}$,
  while the much more rapid decays of the normal 
  outbursts exhibit a faster-than-exponential shape.
    Our modeling using the basic accretion disk limit cycle
  can produce the main features of the V344 Lyr light 
  curve, including the
 peak outburst brightness. 
 Nevertheless there are obvious deficiencies in
    our model light curves:
 (1) The rise times we calculate,
  both for the normal and superoutbursts,
   are too fast.
  (2) The superoutbursts are too short.
  (3) The shoulders on the rise to superoutburst
 have more structure than the shoulder 
            in the observed superoutburst
     and are too slow, comprising about a third to half of the total viscous
  plateau, rather than the $\sim$10\% observed.
   However, one of the
     $\alpha_{\rm cold} \leftrightarrow \alpha_{\rm hot}$
    interpolation schemes
   we investigate (one that is physically motivated)
   does yield longer superoutbursts
  with suitably short, less structured shoulders.
\end{abstract}

\vfil\eject
\section{Introduction}

Dwarf novae
  constitute a subclass of the 
cataclysmic 
    variables (Warner 1995b),
   semi-detached 
   interacting binaries in which a Roche-lobe
filling secondary transfers matter to a more massive  
                       white dwarf (WD).
  The transferred material possesses angular momentum,
  and therefore can only ``fall'' toward the primary down to a 
  radius determined by its specific angular momentum 
  (Lubow \& Shu 1975). This angular momentum barrier is overcome
   by angular momentum transport within the accretion disk
  which carries angular momentum outward and mass inward.
   Current thinking about 
  the physical mechanism responsible for
accretion centers on
   the magneto-rotational instability ($=$MRI; Balbus \& Hawley 1998)
in which the shearing amplification of a weak seed magnetic field leads
to turbulent transport.

V344 Lyr is a short period dwarf nova, of
  subtype SU UMa ($P_{\rm orb}=2.1$ h; Still et al. 2010)
   exhibiting a supercycle pattern (109.6 d; Kato 
et al. 2002) consisting of superoutbursts
   separated by normal outbursts.
 Superoutbursts are long outbursts with slow,
 exponential decays exhibiting superhumps $-$
  modulations in the light curve at a period greater
than the orbital period by a few percent.
  The data presented
  in Kato et al. reveal  only the brightest portions
  of the superoutbursts; the fainter portions 
and all of
    the normal outbursts lie below the lower limit of detection.

  The {\it Kepler} data set for 
                     V344 Lyrae ({\it Kepler} ID 7659570)
$-$ 270 d at 1 min cadence $-$
   provides an extraordinary resource for
accretion disk modelers. The accretion disk limit cycle mechanism,
which is currently employed to account for dwarf nova outbursts,
is based on the storage of material during quiescence in a 
non-steady-state configuration, followed by a dumping of matter onto
the accreting WD when a critical surface density is attained in 
the disk, producing an outburst. During outburst the disk is roughly
in steady state, except for the action of a cooling front that begins
at the outer disk and moves inward.
 The thermal-viscous limit cycle model is 
   not a complete model for the accretion
  disk, but rather primarily 
   a model for the outbursts;
  for instance
    it fails to account 
  for the high X-ray luminosities
  seen in dwarf novae in quiescence
   (Wheatley, Mauche, \& Mattei 2003,
    Mukai, Zietsman, \& Still
    2009).  

  Previous attempts at constraining physical parameters of the model
have relied on AAVSO (American Association of Variable Star Observers)
   and  RASNZ (Royal Astronomical Society of New Zealand)
    data for the long term light curves. 
These data are typically given as 1d means, with large attendant
error bars ($\sim$0.3$-$0.5 mag). The {\it Kepler} data provide an entirely new level
of precision, and   allow      detailed model  constraints. The viscosity
parameter $\alpha$ (Shakura \& Sunyaev 1973) 
           sets the disk time scales. To first order, $\alpha_{\rm cold}$
controls the recurrence time for outbursts, and $\alpha_{\rm hot}$ sets the 
duration of the outbursts. Previous efforts, for instance by Smak (1984$=$S84),
utilized the ``Bailey relation'' relating the rate of the fast decay in 
DNe with orbital period to infer that $\alpha_{\rm hot} \simeq 0.1-0.2$. This remains
the only firm constraint we have in astronomy for $\alpha$ in ionized gas,
and has served as a benchmark value for  modelers calculating the
efficiency of the MRI in accretion disks
(e.g., Hirose et al. 2009).
   By comparing theoretical modeling of the MRI with the observational
  constraints, King, Pringle, \& Livio (2007) showed that theoretical
   values tend to be at least an order of magnitude smaller.
   Most of the simulations have studied the growth 
  and subsequent nonlinear saturation of the MRI using
  idealized ``shearing boxes'' without realistic vertical structure,
    and for such calculations the asymptotic strength of the magnetic
  field decreases with increasing resolution 
 (Fromang \& Papaloizou 2007).
  Vishniac (2009)  argues that current models  are not sophisticated 
  enough       in terms of their dynamic range and treatment of the
 vertical disk structure  to allow a realistic treatment of such important
   effects as magnetic buoyancy and vertical transport of magnetic
  helicity by turbulent eddies,
   and therefore comparisons of theoretical $\alpha$ values with those
   inferred from observations are specious at the present time.

Our goal in this work is to find model parameters that can
account for the gross properties of the long term V344 Lyr light curve.
  There will be additional features in the light curve that can be 
compared to the detailed model light curves in order to gain insight
into shortcomings of the model. We also investigate systematic effects
such as the number of grid points and time step size.

In \S 2 we present an overview of the physics of accretion disks.
In \S 3 we discuss   the V344 Lyr light curve,
in \S 4 we present a review of viscous decays in
              several well-studied dwarf novae, 
in \S 5 we detail the numerical model,
in \S 6 we present the results,
in \S 7 we discuss the results in a larger context,
 and in \S 8 we sum up.

\section{Accretion Disk Physics}

\subsection{Axisymmetric Models}

In the accretion disk limit cycle
model, gas accumulates in quiescence and accretes onto
the central object in outburst 
    (e.g., Cannizzo 1993a, Lasota 2001 for reviews).
  The phases of quiescence and outburst
 are mediated by the action of heating and cooling fronts
that traverse   the disk and bring about phase transitions
   between low and high states, consisting of neutral and ionized
gas, respectively.
    During quiescence, when the surface density $\Sigma(r)$ 
   at some  radius within the disk exceeds a critical value $\Sigma_{\rm max}(r)$,
      a transition  to the high state is initiated;
    during outburst, when $\Sigma(r)$ 
   drops below a different critical value $\Sigma_{\rm min}(r)$,
      a transition  to the low state is initiated.
Low-to-high transitions can begin at any radius, 
whereas high-to-low transitions begin at the outer disk edge.
   This situation comes about because in the outburst disk $\Sigma(r) \propto r^{-3/4}$
(roughly), and the critical surface densities both increase with radius.
   Since the disk mass accumulated in quiescence
  is bounded by $\Sigma_{\rm max}(r)$
   and
    $\Sigma_{\rm min}(r)$,
  one can define a maximum disk mass
\begin{equation}
 M_{\rm disk,  max} = \int 2\pi r dr \Sigma_{\rm max}(r)
\end{equation}
  and a 
  minimum disk mass
\begin{equation}
 M_{\rm disk,  min} = \int 2\pi r dr \Sigma_{\rm min}(r)
\end{equation}
  which will bound the general, time dependent disk mass.
%

   For normal, ``short'' outbursts,
    only a few percent of the stored gas accretes
  onto the central object:
         the thermal time scale of
 a thin disk is short compared to the viscous time scale,
  and the cooling front that is launched from the outer  
  edge of the disk almost as soon as the disk  enters  into outburst
    traverses the disk and reverts it back to quiescence.
   For disks that have been ``filled'' to a higher level with 
respect to $M_{\rm disk,  max}$, the surface density
   in the outer disk can significantly exceed the critical surface
density $\Sigma_{\rm min}$.
     In order for the cooling front to begin,
      however,
       the outer surface density $\Sigma(r_{\rm outer})$
   must drop below 
    $\Sigma_{\rm min}(r_{\rm outer})$.
   Disks in this state generate much longer outbursts, with slower ``viscous''
plateaus, because the entire disk must remain in its high, completely ionized
state until enough mass has been lost onto the WD for the condition 
    $\Sigma(r_{\rm outer})  <  \Sigma_{\rm min}(r_{\rm outer})$  to be satisfied.
%
The slow decay associated with  a  superoutburst
is a direct    reflection of the viscous time scale in the outer disk,
whereas the subsequent faster decay reveals the thermal time scale.
Although the decays of dwarf novae at a given orbital period
are fairly uniform, and form the basis for the Bailey relation,
the rise times show    a greater variety, reflecting the fact 
that the outburst can be triggered anywhere in the disk (Cannizzo
1998b).  Inside-out bursts tend to produce slow-rise times, whereas
outside-in bursts produce fast rises (Cannizzo, Wheeler, \& Polidan
1986). 
%
%
%
Van Paradijs (1983) undertook a statistical study of outburst
properties of dwarf novae spanning a range in orbital period, from 
$<2$ hr to $\sim$10 hr.
   By studying separately the behavior of long and short
outbursts versus orbital period, he found that the apparent strong 
contrast between superoutbursts and normal outbursts in the SU UMa
  systems is due to the fact that the widths of the short (narrow)
outbursts scales with orbital period, whereas the widths of the
  long (wide) outbursts is relatively constant.
  Van Paradijs' argument is strengthened by the more extensive
  data set examined  by Ak, Ozkan, \& Mattei (2002)
  in which the linear trend in the narrow width-$P_{\rm orb}$
  relation is more obvious 
(compare
  Fig. 4 from Ak et al. with Fig. 3b from van Paradijs).
 In these studies the widths of the outbursts were ``corrected''
   by subtracting the rise and decay times.

   For any dynamical system in which a variety of timescales
   contribute to a physical observable, in this case the flux 
from the accretion disk, the controlling time scale will be the slowest
one. Since all the disk timescales $-$  dynamical, thermal, and viscous $-$
  increase with radius, the time scale associated with the viscous
 plateau provides direct information about the viscous time scale $t_{\rm visc}$
   at the outer edge of the disk.

  Warner (1995ab) presents a simple physical model for the
state of the disk during superoutburst.
   During the superoutburst the disk is roughly in steady state
with $\Sigma(r)\propto r^{-3/4}$ out to some $r_d$ which represents
   the radius at which the outward moving heating front stalls.
  The rate of accretion within the disk is much greater than
  the mass transfer rate from the secondary star, which
 can be neglected during this time.
  Therefore the mass of the hot, accreting disk is
\begin{equation}
M_{\rm disk} \approx {8\pi\over 3} r_d^2 \Sigma_{\rm min}(r_d).
\label{brian1}
\end{equation}
  From the standard accretion disk theory relating  
   effective temperature with accretion rate in a 
   steady-state disk (Shakura \& Sunyaev 1973),
    one has
\begin{equation}
{\dot M}_{\rm disk}  = - {8\pi\sigma\over 3GM_1} T_{\rm eff}^4(r_d)  r_d^3,
\label{brian2}
\end{equation}
   where $T_{\rm eff}(r_d) = T_{\rm eff}(\Sigma_{\rm min}[r_d])$ 
  is the effective temperature associated with $\Sigma_{\rm min}(r_d)$.
    Combining eqns. [\ref{brian1}] and [\ref{brian2}] 
(same as eqns. [15] and [16] from Warner 1995a,
       and [3.46] and [3.47] from Warner 1995b) 
       yields
\begin{equation}
{dM_{\rm disk} \over M_{\rm disk}} \approx  - {5 r_d \sigma T_{\rm eff}^4(r_d) 
                \over 3GM_1 \Sigma_{\rm min}(r_d)}  dt.
\label{brian3}
\end{equation}
   Since the right hand side (excluding $dt$)
    is constant, the solution for the 
mass of the hot disk, and also its derivative, is a
  decreasing exponential function.
  Using power law scalings for $T_{\rm eff}(r_d)$ and $\Sigma_{\rm min}(r_d)$,
   setting $r_d$ equal to the 3:1 resonance radius with the 
 binary orbital period, and noting  that luminosity varies as ${\dot M}_{\rm disk}$,
   one  can show (cf. eqn. [20] of Warner 1995a)
\begin{equation}
 t_{\rm visc}  \approx  17 \ {\rm d} 
               \ \alpha_{-1}^{-4/5}
               \ P_h^{1/4}
               \  m_1^{1/6},
\label{brian4}
\end{equation}
  where $\alpha_{-1}=\alpha/0.1$,
   $P_h$ is the orbital period in hr,
   and $m_1=M_1/\msun$.
  (The $\alpha$ in this context is the hot state value.)
  Within the
viscous plateau portion of a superoutburst,
   the instigation and propagation of a
cooling front at the outer edge
   is thwarted
by virtue of excessive mass in the disk.
   This phenomenon is not new and has been seen 
  in numerous previous  studies
  (e.g., 
   Cannizzo 1993b$=$C93b,
  Hameury et al. 1998$=$H98,
   Ludwig \& Meyer 1998,
    Buat-M\'enard, Hameury, \& Lasota 2001,
  Schreiber, Hameury, \& Lasota 2003).
   The viscous time does not represent
the duration of the outburst itself, but rather the 
  time scale (approximately in d mag$^{-1}$) associated with the
  viscous plateau;
  it represents an $e-$folding time for the decrease 
  of disk mass.

\subsection{Non-axisymmetric Models}

Whitehurst (1988) discovered a  hydrodynamical instability 
that can occur in the outer parts of accretion disks
when the mass ratio $q= M_2/M_1$  is less  than  1/4.
    For these extreme mass ratio systems,
expected  in CVs below the period gap if the secondary fills
its Roche lobe, the outer part of the disk can lie beyond
the point of 3:1 resonance with the binary period.
The outer disk can then be excited by the inner
Lindblad resonance (Lubow 1991ab), causing the global disk 
oscillation mode that is observed as common
superhumps. Before the onset of the superhump oscillation, there 
are well-known spiral dissipation waves in the
disk that are fixed in the co-rotating frame. Once the superhump 
oscillation begins, these spiral arms advance
$180^\circ$ every superhump cycle\footnote{See, for example,
   www.astro.fit.edu/wood/visualizations.html.}. As
a given arm passes between the two stars, it expands outward in 
the shallower potential, but as fluid in that
arm compresses back into the disk as the disk is maximally 
distorted, viscous dissipation causes a peak in the
photometric light curve (Simpson \& Wood 1998).
 A secondary source of the superhump signal is caused by the
variable distance the mass stream must fall from the 
 L1 point to the edge of the non-axisymmetric, flexing
disk.
  Because the outer disk is thought to expand
during superoutburst, potentially beyond the 3:1 radius,
    the  ``tidal instability''  model
  provides a natural explanation for superhumps.  If the mass ratio
is extreme  enough,  superhumps can also occur outside of superoutbursts
  because the disk can remain  extended beyond the 3:1 radius,
  and therefore continue oscillating  even  into quiescence.
  Osaki (1989ab) combined the accretion disk limit cycle model
with Whitehurst's tidal  instability model and  proposed
the ``thermal-tidal'' instability model (TTIM) for superoutbursts.
   This  model  posits that the presence of an oscillating
precessing disk, triggered by the migration of gas beyond the 3:1 resonance
radius,  also leads to a greatly  enhanced tidal torque  
  acting on the
disk. This ultimately  contracts the  outer disk and leads to
a superoutburst.  Thus in the TTIM  one has the usual
   hysteresis relation between surface density and
temperature in the disk,  and also a hysteresis  relation
  between the total angular momentum of the disk
and  the   tidal  torque acting on the outer disk.
(The latter hysteresis is somewhat speculative.)
   This model was examined in detail
   and compared to the enhanced mass transfer model (EMTM)
   for superoutbursts  by
   Ichikawa, Hirose, \& Osaki (1993) and later workers, 
most recently  Schreiber et al. (2004).

For completeness,  we note that
               Smak has recently  challenged 
the conventional wisdom that superhumps
 are due to a precessing, eccentric disk (Smak 2009abcd).
  Smak (2009a) presents  a re-analysis of data
  taken by earlier workers on Z Cha,
 WZ Sge, OY Car, and IY UMa and claims that a variety of
   errors, including miscalculated beat phases and
 an assumption that observed eclipses are pure disk
eclipses, led to erroneously large values of the
eccentricity  $e$.
 Smak's re-analysis of the disk eccentricity in Z Cha
   leads to $e=0.05\pm0.05$.
 Smak (2009b) compares in detail observed values for the
amplitudes of superhumps with theoretical values
   determined from bolometric light curves produced
 by smooth-particle hydrodynamics calculations. He finds
the observed values to be $\sim$10 times larger than the 
theoretical values.
  Smak (2009c) presents a re-analysis of five eclipses
  of Z Cha observed by Warner \& O'Donoghue (1988).
  Smak  examines two local minima at orbital phases
  $\Phi\approx -0.05$ and $0.04$.  He argues that the
  first minimum is  not due to an occultation but rather
absorption in the mass overflow stream, and the second
minimum coincides with the trajectory of the mass overflow stream.
The  upshot of his analysis is that superhumps
  are due to modulated mass transfer which leads to
periodically enhanced dissipation of the stream  kinetic energy.
  Smak (2009d) presents evidence  for periodically variable 
  irradiation of secondary components  
   and proposes that superhumps are due to enhanced 
 dissipation of the stream  kinetic energy.
     He outlines specific future modeling efforts that might
clarify the overall picture.

\section{ {\it Kepler} light curve of V344 Lyr}

The bandpass utilized by {\it Kepler} is
   considerably broader than the standard 
Johnson $V$ band. Figure 1 gives a
  comparison of the two filters
\footnote{The {\it Kepler} response is taken from http://keplergo.arc.nasa.gov/CalibrationResponse.shtml.}.
Throughout this paper we will 
  refer to 
 Johnson $V$ apparent and absolute magnitudes
   using $V$ and $M_V$,
  and
 {\it Kepler} apparent and absolute magnitudes
  using $Kp$ and $M_{\rm Kp}$.
Figure    2         shows the {\it Kepler} light curve
 of V344 Lyr.
  The observation window
  contains 19
    outbursts showing
  the sequencing SSSLSSSSSSSLSSSSSSS, where S$=$short outburst and L$=$long
 outburst, i.e., a superoutburst. 
%
Based on photographic observations made by Hoffmeister (1966), Kato (1993)
estimated that the cycle
length for normal outbursts was 16$\pm$3 d. An estimate of this 
cycle length prior to {\it Kepler} was difficult
   because
   the normal outbursts of V344 Lyr are short-lived and reach only 
to near $Kp=16$ (or $V\simeq17$). 
   In Table 1 we list the durations (using $Kp=18$ as the cut line)
   and amplitudes (measured from the local
  mean quiescence level)
    of the 19 observed outbursts (17 normal and two superoutbursts).
  Both the recurrence
 times and amplitudes of the normal outbursts
  appear to reach a maximum
 roughly midway between superoutbursts, rather than 
increasing monotonically between consecutive superoutbursts as in the TTIM.
           We also note that the level of quiescence is $\sim$1 mag 
higher after the superoutburst, and declines with an $e-$folding time
 of $\sim50$ d.
   This trend may be associated with the cooling of the WD 
  after being heated by the mass accreted during superoutburst
  (e.g., Sion 1995).


The superoutbursts last   $\sim17$ d and 
  reach   a maximum amplitude 
$\Delta Kp = 4.4-4.5$ mag.
   The decline of the superoutbursts are  close to  linear
 (plotted as mag versus time, 
    or exponential as flux versus time) 
  for $\sim 11$ d,
  after which they  fall
rapidly ($\sim$2 d) back to quiescence. During this part
the decline rate is 
0.083 mag d$^{-1}$, in good agreement 
  with the value of 0.094 mag d$^{-1}$ found 
by Kato (1993).
%
The amount of time spent at $Kp\la 16$
  is $\sim$14 d,
  and the duration of the entire plateau
        is $\sim$12 d.
The 
      decay rate
   is 
  $\sim$12 d mag$^{-1}$,
  therefore the plateau portion of the decay
encompasses $\sim$1  mag  $-$ a dynamic range of $\sim$2.5, 
  or slightly less than one $e-$folding ($\sim$2.7$\times$).
  There are  also   $\sim$0.5 mag shoulders
  on the superoutburst rises lasting $\sim$1 d,
  which appear to be normal outbursts embedded at the beginning 
  of the superoutbursts.

 It is noteworthy that the fast decays
  of the normal outbursts and superoutburst
have a faster-than-exponential shape
(i.e., concave downward when 
   plotted as magnitude versus time).
%
S84 utilized the Bailey relation between  the fast rate  of decay
 in dwarf novae and their orbital period to constrain $\alpha_{\rm hot}\simeq 0.1-0.2$.
   The large errors associated with the AAVSO data were consistent 
with exponential decay.  
   Cannizzo (1994) found                  that  to account for the 
supposedly exponential decays,  which   correspond to the  time during
which a cooling front traverses the disk, $\alpha$ must vary weakly with radius
($\simpropto    r^{0.3}$). 
   The faster-than-exponential decays
   captured by {\it Kepler} in V344 Lyr
 negate  this result and  indicate  that a constant $\alpha$
  scaling  with  radius is  consistent with the data.

\section{Viscous Plateaus in Dwarf Novae}


   The Bailey relation between the fast rate of decay of dwarf nova
 outbursts and orbital period 
   corresponds 
   to the time interval 
 during which the cooling front traverses the disk.
   Some long dwarf nova outbursts  have enough dynamic range
   in $V$ that one can also use the slow decay portion, the viscous
 plateau, to infer an $\alpha$ value. These inferred values 
   agree with those determined from the Bailey relation 
 (Warner 1995ab; 
 Cannizzo 2001b, Cannizzo et al. 2002).

The decay rates associated with several dwarf novae exhibiting
  viscous plateaus have been fairly well established, and enable
  one to make a reasonable estimate of the decay rate 
  expected in a system with the orbital period of V344 Lyr.  
 The {\it Kepler}  observation gives 
   a decay rate within the superoutburst 
  $\sim$12 d mag$^{-1}$.
  We associate this with
  the viscous time,
  which is basically an $e-$folding time for a
  perturbation to the $\Sigma(r)$ profile to be
  damped out.  For a disk with a  $\Sigma(r)$ profile
  approximating steady state, i.e., $\Sigma(r)\propto r^{-3/4}$,
  and for which the mass flow within the disk
  and onto the WD
  greatly exceeds
  the rate of mass addition feeding into the outer disk from the 
  secondary star,
  the viscous time  represents roughly a time for the mass of 
  the disk to decrease by a factor $e$.

   Thus for a given viscous plateau-type outburst, the ratio 
   $t_{\rm plateau}/t_{\rm visc}$  represents the number of $e-$foldings
   by which the mass of the disk decreases due to accretion
   onto the central object.
   It is instructive to look at examples
   of viscous decays in several well-studied
   dwarf novae:



\noindent{\bf SS Cyg\footnote{
  If one accepts the {\it HST}/FGS trigonometric parallax
  for SS Cyg $6.02\pm 0.46$ mas
    (Harrison et al. 1999) which implies $D=166\pm13$ pc,
  the accretion disk limit cycle model is not able to reproduce
  dwarf nova outbursts for SS Cyg 
(Schreiber \& Lasota 2007, Smak 2010).
%
%
%
  Harrison et al. (2004)
  acquire           spectra of  MK standard stars
  to refine the mean absolute parallax of the reference frame
  and present a reanalysis of the SS Cyg parallax (see their sect. 2.3).
  They 
  obtain a revised value $6.06\pm 0.44$ mas $-$ close to their original value.
%
%
}:}
Long outbursts in SS Cyg ($P_{\rm orb}=6.6$ hr)
             have a duration of $\sim$10 d (Cannizzo \& Mattei 1992).
   and the viscous time in the outer disk $t_{\rm visc} \approx 40$ d.
  Thus $t_{\rm plateau} \ll t_{\rm visc}$,
  and only about 20\% of the disk mass is accreted onto the WD (C93b).
 
\noindent{\bf U Gem:}
In the 1985 long outburst of U Gem ($P_{\rm orb}=4.25$ hr), 
   the decay
rate during the viscous plateau was $\sim$26
  d mag$^{-1}$ over the $\sim$35 d of the burst (Cannizzo et al. 2002).
   Since $t_{\rm visc} \approx t_{\rm plateau}$,
             $\sim$70\% of the disk mass accreted onto the WD.

\noindent{\bf V344 Lyr:}
In the superoutbursts of V344 Lyr ($P_{\rm orb}=2.1$ hr)
   shown in Fig. 2
   the locally defined decay
rate during the plateau remains  constant at $\sim$12
  d mag$^{-1}$  over most of the $\sim$14 d of the burst,
  thus as with  U Gem
    $\sim$70\% of the disk mass accreted onto the WD.

\noindent{\bf WZ Sge:}
In the 2001 superoutburst of WZ Sge ($P_{\rm orb}=81$ min), 
   the locally defined decay
rate during the plateau increased from $\sim$4 to $\sim$12
  d mag$^{-1}$  over the $\sim$20 d of the burst (Cannizzo 2001b).
  The fact that  $t_{\rm plateau} \approx 3 t_{\rm visc}$
   means that only a few percent of the initial disk mass
   remained at the end of the superoutburst.
   (An alternative considered in the Discussion is that 
  irradiation-induced enhanced mass overflow from the secondary
  star prolonged the superoutburst.)
  Without a very low $\alpha_{\rm cold}\simeq10^{-5}$
  there is not enough accumulated mass in the disk
to account for the outburst's fluence (Smak 1993) and the
 long recurrence time.

In summary,  one can readily find examples of known systems with viscous 
type outbursts in which the duration of the outburst itself is either 
greater than, less than, or about the same as the viscous time scale
evaluated at the outer edge of the accretion disk.  The
   actual course
   taken by 
   a given system depends on the degree of overfilling of
the disk in quiescence with respect to $M_{\rm disk, min}$.



\section{Numerical Modeling}

  Our numerical model is discussed extensively in C93b
  and subsequent papers.
    The basic strategy is the time dependent solution of two
  coupled differential equations, one giving the $\Sigma(r)$
  evolution, and one the evolution of disk midplane temperature $T(r)$.
  Thus the disk is entirely unconstrained as
  regards
   deviations from steady state and thermal equilibrium.
      Many previous workers have presented time dependent
calculations by solving these equations, or similar versions
 (e.g.,
  S84; 
   Lin, Papaloizou, \& Faulkner 1985$=$L85; 
   Mineshige \& Osaki 1983$=$M83, 1985$=$M85, 
   Mineshige 1986$=$M86, 1987$=$M87;
   Pringle, Verbunt \& Wade 1986$=$P86;
   Cannizzo et al. 1986$=$C86, C93b, 1994, 1998a, 2001a;
   Angelini \& Verbunt 1989;
   Ichikawa \& Osaki 1992;
   Meyer \& Meyer-Hofmeister 1984,
   Ludwig \& Meyer 1998;
                         H98, 
   Menou, Hameury, \& Stehle 1999,
   Buat-M\'enard, Hameury, \& Lasota 2001,
   Buat-M\'enard, \& Hameury 2002,
   Schreiber, Hameury, \& Lasota 2003, 2004$=$S04).

   Improvements in our  code
   made since the original version include the
provision for a  variable outer disk radius, self-irradiation,
  and evaporation with various possible laws
     (e.g., Meyer \& Meyer-Hofmeister 1994).
The diffusion equation governing the evolution
of surface density is given by (Lightman 1974, Pringle 1981)
\begin{equation}
{\partial  \Sigma  \over \partial t} =
{3\over r} {\partial  \over \partial r}
\left[ r^{1/2} {\partial  \over \partial r} 
 \left( \nu \Sigma r^{1/2} \right)\right],
\label{viscous_evolution}
\end{equation}
where $\Sigma(r,t)$ is the surface density 
and $\nu$ is the kinematic viscosity coefficient,
%
%
\begin{equation}
  \nu  = {2\over 3}  {\alpha \over \Omega} { {\mathcal R} T \over \mu}.
\end{equation}
We follow the technique of Bath \& Pringle (1981) 
  in discretizing the $\Sigma(r,t)$ evolution
  equation.

The thermal energy equation
  governing the evolution of the midplane
disk temperature $T(r,t)$ is given by
%
%
%
%
%
\begin{equation}
   {\partial T\over \partial t} = {2(A-B+C+D)\over {c_p \Sigma }} 
    - {{\mathcal R} T \over {\mu c_p}} {1\over r} {\partial \over \partial r} (r v_r)
    -  v_r  {\partial T \over \partial r},
\label{thermal}
\end{equation}
where the viscous heating
\begin{equation}
    A   =  {9\over 8}  \nu  \Omega^2  \Sigma,
\end{equation}
the radiative cooling
\begin{equation}
   B    = \sigma T_e^4,
\end{equation}
\begin{equation}
     C =  {3 \over 2} 
  {1\over r}  {\partial \over \partial r} 
  \left(  c_p \nu  \Sigma  
   r  {\partial T\over \partial r}
    \right),
\end{equation}
and
\begin{equation}
  D =  {h\over r} {\partial \over \partial r} 
\left(r {4acT^3\over 3\kappa_R\rho} 
{\partial T \over \partial r} \right).
\end{equation}


 Of the four terms appearing in the numerator of the first
term of thermal energy equation, the first two $A$ and $B$
  are fairly standard.
   The term $C$ represents the radial heating 
flux due to turbulent transport, and $D$ is that due to 
radiative transport. 
 Previous workers have used varying forms of $C$ and $D$.
 Some early studies adopted $C=D=0$ (M83), $C=0$ (L85), or 
  a form of $C$ 
   not
  expressible as
    the divergence of a flux
  (M83, M86, C93b; 
   see H98 for a critical discussion).

The local flow velocity $v_r$, which
 can vary enormously
   from inflow to outflow
   over the radial
extent of the disk, depending on whether or not 
transition fronts are present,
  is given by
\begin{equation}
 v_r = - {3\over \Sigma r^{1/2}} {\partial \over \partial r} 
\left(\nu\Sigma r^{1/2}\right),
\end{equation}
%
where upwind differencing is implemented 
   over the radial grid as follows:

\begin{eqnarray}
   s_a & = & \nu_{i-1} \Sigma_{i-1} {r_{i-1}}^{1/2} \nonumber \\
   s_b & = & \nu_{i  } \Sigma_{i  } {r_{i  }}^{1/2} \nonumber \\
   s_c & = & \nu_{i+1} \Sigma_{i+1} {r_{i+1}}^{1/2}  
\end{eqnarray}

\begin{eqnarray}
   q_{-} & = & - {3 \over {\Sigma_i {r_i}^{1/2}}}  { {s_b-s_a} \over {r_i     - r_{i-1}}} \nonumber \\
   q_{+} & = & - {3 \over {\Sigma_i {r_i}^{1/2}}}  { {s_c-s_b} \over {r_{i+1} - r_{i  }}}  
\end{eqnarray}

\begin{eqnarray}
   {\rm if}\left( {q_{-}\over q_{+}} > 0 \ \ {\rm  and  } \ \ q_{-} > 0 \right) &  \ \ \ \ \ \ \ (v_r)_i = & \max(q_{-}, q_{+} ) \nonumber \\ 
   {\rm if}\left( {q_{-}\over q_{+}} > 0 \ \ {\rm  and  } \ \ q_{-} < 0 \right) &  \ \ \ \ \ \ \ (v_r)_i = & \min(q_{-}, q_{+} ) 
\end{eqnarray}

\begin{eqnarray}
   {\rm if}\left( {q_{-}\over q_{+}} < 0 \ \ {\rm and} \ \ {{|q_{-}|}\over{|q_{+}|}}>1\right) & \ \ \ \ \ \ \ \ \ \ (v_r)_i  = &  q_{-} \nonumber \\
   {\rm if}\left( {q_{-}\over q_{+}} < 0 \ \ {\rm and} \ \ {{|q_{-}|}\over{|q_{+}|}}<1\right) & \ \ \ \ \ \ \ \ \ \ (v_r)_i  = &  q_{+}. 
\end{eqnarray}

The  scalings for the local maximum and minimum in the $\Sigma - T$ relation
(from Meyer \& Meyer-Hofmeister 1981, 1982) are
\begin{equation}
\Sigma_{\rm max} = 444 \ {\rm g} \ {\rm cm}^{-2} \ r_{10}^{1.1} m_1^{-0.37} \alpha_{c,-2}^{-0.7},
\end{equation}
where $\alpha_{c,-2}= \alpha_{\rm cold}/0.01$,
and
\begin{equation}
\Sigma_{\rm min} = 44.4 \ {\rm g} \ {\rm cm}^{-2} \ r_{10}^{1.1} m_1^{-0.37} \alpha_{h,-1}^{-0.7},
\end{equation}
where $\alpha_{h,-1}= \alpha_{\rm hot}/0.1$.
  The disk midplane temperatures associated with
these extrema are
\begin{equation}
T(\Sigma_{\rm max})  =  5275 \ {\rm K} \                          \alpha_{c,-2}^{-0.3},
\end{equation}
and 
\begin{equation}
T(\Sigma_{\rm min})  =  35900 \ {\rm K} \ r_{10}^{0.064} m_1^{-0.02} \alpha_{h,-1}^{-0.16}.
\end{equation}
 The equilibrium temperature scalings
  for the stable branches are
   given in C93b (taken from Cannizzo \& Wheeler 1984).
%
%
%
During transitions when the local disk temperature
 lies between the values associated with the local
maximum and minimum in surface density,
  $T[\Sigma_{\rm max}(r)] < T(r) < T[\Sigma_{\rm min}(r)]$,
we use a logarithmic interpolation
 to obtain the local $\alpha$ value,
\begin{equation}
{\log}_{10}\alpha(r) = {\log}_{10}\alpha_{\rm cold} + f, 
\end{equation}
where
\begin{equation}
  f =   { { T -  T(\Sigma_{\rm max}) }
            \over 
        {T(\Sigma_{\rm min}) 
      -  T(\Sigma_{\rm max}) }}
       ( {\log}_{10} \alpha_{\rm hot} - {\log}_{10} \alpha_{\rm cold} ).
\label{log}
\end{equation}
%
%
%
   In tests we also consider
 a logarithmic interpolation factor $f$
introduced by H98,
\begin{equation}
  f = \left[ 1 + \left( {T_0\over T} \right)^8  \right]^{-1}
       ( {\log}_{10} \alpha_{\rm hot} - {\log}_{10} \alpha_{\rm cold} ),
\label{hameury}
\end{equation}
as well as a linear interpolation
\begin{equation}
  \alpha(r) = \alpha_{\rm cold}
          + 
     { { T -  T(\Sigma_{\rm max}) }
            \over 
        {T(\Sigma_{\rm min}) 
      -  T(\Sigma_{\rm max}) }}
       ( \alpha_{\rm hot} - \alpha_{\rm cold} ).
\label{lin}
\end{equation}


In calculating the effects of the tidal torque from the
secondary we  follow Smak (1984) and later workers
  who used a formalism 
   developed by Papaloizou \& Pringle (1977)
   in which the tidal torque 
   varies as the fifth power of radius.
  This leads to a depletion in disk material
\begin{equation}
    {\partial\Sigma \over \partial t}  = 
   -  c_1 \omega 
    {   \nu \Sigma
   \over  {2 \pi    j_s} }
   \left( r \over a \right)^5,
\end{equation}
where $\omega$ is the orbital
  angular frequency $2\pi/P_{\rm orb}$,
  the orbital separation is $a$,
 and the specific angular momentum $j_s = (G M_1 r)^{1/2}$.
   We do not reset the value of $c_1$
  during the course of a run
    as in the TTIM.
   Ichikawa \& Osaki (1994) presented detailed calculations of the strength
 of the tidal torque versus distance at large radii within the disk
and found that the power law ($\propto r^5$) determined by Papaloizou \& Pringle
  is only valid at one radial location in the outer disk; further out
  the torque relation   steepens
considerably, approaching infinite slope
  at the last non-intersecting orbit.
   Therefore 
   in some sense the more naive ``brick wall'' condition
for the outer edge used by C93b
 and other workers might be more descriptive. 
   Regardless of which prescription is used,
   if the parameters associated with the outer edge are not 
  manipulated during the run, the detailed treatment is not important
to the overall model. Using a different end point treatment would 
result in slightly different optimal parameters.

  Given a disk temperature $T$ and surface density $\Sigma$,
  the disk semithickness $h$ and density $\rho$
   are determined from vertical hydrostatic equilibrium
   to yield
\begin{equation}
    h  = { {P_r + \sqrt{ P_r^2 + c_2 c_3 }}  \over c_2},
\end{equation}
  where
\begin{equation}
       c_2   = \Sigma  \Omega^2,
\end{equation}
\begin{equation}
      c_3   = \Sigma  { {{\mathcal R}  T}\over \mu},
\end{equation}
  the radiation pressure
\begin{equation}
    P_r   = {1\over 3}  a T^4,
\end{equation}
  and the density
\begin{equation}
   \rho  = {\Sigma \over {2 h}}. 
\end{equation}

  To determine the computational time step, we take
\begin{equation}
\Delta t = f_t \min\left[ {\min}_i\left(\partial\ln \Sigma_i \over \partial t\right)^{-1},  \ \ 
                          {\min}_i\left(\partial\ln    T_i   \over \partial t\right)^{-1} \right], 
\end{equation}
  where $\min_i$  refers to
  the minimum over all grid points $i$,
  and $f_t$ is a small number between $1/80$ and $1/640$.
  The quantities
     $\partial\ln \Sigma_i/\partial t$ and
     $\partial\ln    T_i/\partial t$ are evaluated
  using the evolution equations.

\section{Results}

 To calculate the long-term light curves
   from our numerical trials,
  we assume local  Planckian 
   flux distributions
  for each annulus, and
in each time step integrate
 over the radial profiles
of effective temperature.
  We compute
 the absolute {\it Kepler} magnitude $M_{\rm Kp}$, 
  using the filter shown in Fig. 1,  assuming a
  face-on disk. 
   To account for the background
   contribution, assumed constant,
 we add the secondary star and hot spot.
For the secondary, we assume an M5V star with radius
   $0.221R_\odot = 1.54\times 10^{10}$ cm and $T_{\rm eff}=2951$ K.
This contributes $M_{\rm Kp}=11.31$.
  For the hot spot, assuming a 12000 K emitting circular
region with radius $10^9$ cm gives 
  $M_{\rm Kp}=10.41$.
  As a test of our magnitude subroutine, using solar parameters
$R_\odot=6.96\times 10^{10}$ cm and $T_{\rm eff}=5778$ K
we find $M_V=4.85$ (the accepted value is 4.83)
  and $M_{\rm Kp}=3.83$.
 (The $M_V$ values for our putative secondary and
  hot spot are 12.97 and 11.30, respectively.)
  A reasonable inclination $i\simeq 45^{\circ}$
 would dim the disk by $\sim$0.4 mag.

  After some experimentation, we
   found a set of parameters
  that reproduces
   approximately the overall
  pattern of supercycle and short outbursts.
We take a central $0.6\msun$ WD accretor,
  an inner disk radius $r_{\rm in}=2\times 10^{9}$ cm,
   and 
   a maximum  outer disk radius $r_{\rm out}=10^{10}$ cm,
   a mass feeding rate into the outer disk
   ${\dot M}_T = 2 \times 10^{-10}\msunyr$, and 
    $\alpha$ values $\alpha_{\rm cold}=0.0025$ and 
                    $\alpha_{\rm hot}=0.1$.  
  The recurrence time for outbursts varies inversely with ${\dot M}_T$ and
   $\alpha_{\rm cold}$, and the duration of outbursts varies inversely with
   $\alpha_{\rm hot}$  (Cannizzo, Shafter \& Wheeler 1988$=$CSW, C93b). 
    Also, the ratio of the number of short to long outbursts in a long sequence varies in a
complicated way on  these  parameters, and on $r_{\rm in}$ and $r_{\rm out}$.
    The strengths of the 
    dependencies of the various time scales
  $-$ recurrence, burst duration, rise, 
  and decay  $-$
   on the input parameters
    vary according to location within the
  multidimensional (${\dot M}_T$, $\alpha_{\rm cold}$, $\alpha_{\rm hot}$, 
        $r_{\rm in}$, $r_{\rm out}$, $m_1$)  parameter  space (CSW, C93b).

    We use the $\alpha$ interpolation given by eqn. (\ref{log}).
   As regards the inner disk radius,
  the value 
 for  $r_{\rm inner}$ is larger
 than the WD itself;
   if we take $r_{\rm inner} \simeq r_{\rm WD}$,
   we tend to get far fewer short outbursts
 between two successive superoutbursts.
   This hint of a large inner
   radius
    may have some observational support
 in the SU UMa systems  (e.g., Howell et al. 1999).
   The  ratio 
  $\alpha_{\rm cold} / \alpha_{\rm hot} =  40$
   is about 10 times greater than
  commonly used in dwarf nova calculations.
  The mass transfer rate from the secondary  ${\dot M}_T$
  is $\ga2$ times the nominal value 
    $\sim0.8\times10^{-10}\msunyr$
 found from evolutionary calculations 
 of  cataclysmic variables
 below the period gap 
  where the evolution is driven
  predominantly by
  gravitational radiation
(e.g., Kolb, King, \& Ritter 1998;
        Howell, Nelson, \& Rappaport 2001). 
%
%
%
When one takes into account the $\sim$20\% depletion of the disk mass
accompanying each of the short (normal) outbursts,
   however,
 the  effective rate of
accumulation of mass in the disk
  between successive superoutbursts
    decreases from $2\times 10^{-10}\msunyr$
to $\sim$0.3$\times 10^{-10}\msunyr$.
  In our models, $\sim$40$-$60\%  of the disk mass is accreted during 
a  superoutburst.

In using a complicated 
   numerical model, it is 
  important to carry out 
  testing to gain an understanding of
systematic effects:

\subsection{Interpolation between $\alpha_{\rm cold}$ and $\alpha_{\rm hot}$}

  Figure       3            shows the effect of
  using a log versus a linear interpolation
   for the $\alpha$ value between 
$\alpha_{\rm cold}$ and $\alpha_{\rm hot}$.
  The linear interpolation leads to smaller amplitude
 outbursts with little or no quiescent intervals.
   The decays are also much slower, and concave
  upwards when plotted as mag versus time 
  (i.e., with a functional form
    that is slower-than-exponential).  
Figure       4
   shows the effect of using the H98 interpolation.
  H98 adopted the constant $T_0=25000$ K in their scaling;
   we also look at smaller values $18750$ K and $12500$ K. 
  The smaller $T_0$ values produce superoutbursts which have the
recognizable plateau and subsequent fast decay, but for the $T_0=12500$ K
run the fast decays are slower-than-exponential.
  Hameury (2002) compared the two cases  $T_0=24000$ K and  $T_0=8000$ K
   and found little difference between the resultant light curves (see his Fig. 3)
   in his model. 

    Figure      5        depicts 
 the different interpolations
          as well as parameterizations of the 
fractional ionization $\xi$. 
   The MRI is thought to mediate the strength of $\alpha$
   through turbulent transport, and the effect of shearing 
on the weak magnetic field embedded in the gas
   is more efficient as $\xi$ increases.
  The H98 scaling for $T_0=12500$ K does a reasonable job in 
approximating the $\rho=10^{-6}$ g cm$^{-3}$ curve for $\xi$,
  and may be more physical than the other scalings.

\subsection{Number of Grid Points $N$}

 Early  time dependent
 accretion disk limit cycle
   studies
 were concerned mainly with showing the ability 
of the disk, under the conditions of an imposed
limit cycle in each annulus, 
  to undergo collective, global oscillations
in which a major part of the disk  participated.
  The total
   numbers of grid points $N$ was small 
  and
  the comparison with observational data 
   minimal
(e.g., $N=20-25$: S84;
          $N=35$: L85;
       $N=22-44$: M85;
          $N=17$: M86;
          $N=40$: P86;
          $N=44$: C86).
   One can see obvious irregularities in many of the early
light curves which are characteristic of numerical instability.
 C93b undertook the first detailed investigation regarding the number
of grid points, and found that at least several hundred were
required to obtain reliable light curves.
   H98 utilized both uniform and adaptive mesh codes to study
the properties of the transition fronts, and found
  a requirement of  $N\ga800$
   for a uniform-equivalent
 mesh to resolve the fronts
         adequately.

Figure         6         shows  a detail of one superoutburst
     for models with $N=200$ to $1600$ grid points.
 In term of the long term light curves, even with $N>1000$
     the sequencing of long and short outbursts is not
   perfectly stable using our explicit code.
    An implicit code such as that described
   in H98 may be more numerically stable.
  In terms of individual outburst
   profiles,  the rise times we calculate
   for both  the normal and superoutbursts
   are too fast.
   In addition, our outbursts are too bright
  by $\sim$0.5 mag. 
  A nominal inclination $i\simeq 45^{\circ}$ would
 make our disks dimmer by $\sim$0.4 mag,
  bringing them
   close to the level observed in outburst. 
   For comparison we also plot the V344 Lyr data,
  where $D=620$ pc was assumed in converting to absolute
magnitude (found by 
   Ak et al. 2008 using a period-luminosity relation).
  There is a slight increase in superoutburst
duration from $N=200$ to $N=800$ due to the fact
  that more mass accumulates 
  as the supercycle lengthens.
In addition, the shoulder in the light
curve also becomes longer.
 For the highest $N$ run the duration
of the superoutburst is still shorter than that seen in V344 Lyr, 
  while the shoulder is too long.
 The superoutburst calculated using the H98 scaling with 
  $T_0=12500$ K better reproduces the observed duration of the V344 Lyr superoutburst,
  and also has a shorter shoulder with less structure than the other calculations.
   However, it suffers from a rise time that is too short,
  and the fast decay is now slower-than-exponential. 
         Also, the viscous decay rate, which has been optimized
  to the observed value using the eqn. (\ref{log}) scaling,
   is now too fast.

\subsection{Thermal Energy Equation}

Considering the many forms of the thermal energy 
equation used by previous workers, and the potential 
limitations of employing a one-dimensional
   hydrodynamical formalism,
%
   we now look at the effects of switching off
various terms in the thermal energy equation
  in an attempt to ascertain the
relative importance of  each term.
  Figure     7      shows 
the effect of setting equal to zero: 
   (i) $C$, 
  (ii) $D$, 
 (iii) $C$ and $D$, 
  (iv) the first   advective term in eqn. (\ref{thermal}),
       and
   (v) the second  advective term in eqn. (\ref{thermal}).
 In the $C=0$ run one sees a rounded superoutburst
and only three short outbursts
  between two superoutbursts.
The (radiative)  $D$ term  is less influential, 
as the superoutburst
profile is    relatively unchanged 
  from the first panel.
Setting $C=D=0$ produces a single dip
  in the rise to superoutburst, and
a longer viscous plateau (i.e., exponential decay).
The first advective term appears to be of
comparably minimal importance as the $D$ term;
  i.e., the omission of either has a negligible effect.
The second advective term is 
the most influential of all terms 
  examined: without it one has much longer
superoutbursts $-$ about 70\% of the accumulated disk mass is
accreted, which is about the same as in V344 Lyr. 
  However, the viscous plateau still comprises less than half of 
  the superoutburst, rather than the $\ga90$\% seen in V344 Lyr.

\subsection{Detuning of Optimal Parameters}

There is a complicated interplay
  between the different parameters
that enter into the overall results.
  Figure  8 
  shows the effect of varying the inner disk radius.
  For smaller  inner disk radii,
 one has a larger dynamic range in $r_{\rm outer}/r_{\rm inner}$
and as a result there tends to be a more complete
   emptying of the disk during outbursts,
  and therefore less of a tendency to have a series
 of short outbursts leading up to a superoutburst.
   The net interaction 
   among 
   all the model parameters is such that the
   observed  sequencing for V344 Lyr is roughly
   reproduced with $r_{\rm inner}\simeq 2\times 10^9$ cm.
   Figure 9
   shows the effect of varying the outer disk radius.
  For larger $r_{\rm outer}$ values not only does
 the number of short outbursts increase, but
  also the shape of the superoutburst distorts
  significantly from the viscous plateau shape.

\subsection{Relevance of Results to the V344 Lyr light curve}




No one light curve we have shown reproduces all 
 features of the observed light curve shown in Fig. 2.
  For instance, there is an atypically long quiescent period at the beginning
  of the observed light curve.
  The recurrence time for outbursts  varies
                  inversely to varying degrees
                    with $\alpha_{\rm cold}$,  $\alpha_{\rm hot}$,
  and ${\dot M}_T$ (CSW).
  Therefore if any one of these parameters
  were anomalously small, that could delay an outburst.
   Also, the pre-superoutburst normal outbursts are brighter 
  than the post-superoutburst normal outbursts,
   and the pre-superoutburst quiescent level is fainter 
  than the post-superoutburst quiescent level. 
    As noted earlier, the disk instability model is ill-equipped to
make any statement about the quiescent state.  
   However, one possibility might be  
             that  ${\dot M}_T$ were elevated for a while
after the superoutburst; that could lead to crowding of
 the short outbursts, and an increase in the quiescent level.
  There may be other causal relations 
  among $\alpha_{\rm cold}$,  $\alpha_{\rm hot}$,
  and ${\dot M}_T$, which are important.

An analysis of the phase variations in the quiescent-state oscillations 
 in the light curve at the beginning of Fig. 2 light curve show
these are likely to be negative superhumps.  As such they indicate
  accretion onto an accretion disk that is tilted a few
degrees out of the orbital plane and precessing slowly
 in the retrograde direction
 (Barrett et al. 1988;
   Foulkes et al. 2006;
   Wood \& Burke 2007,
  Wood et al. 2000, 2009).
%
Instead of impacting the rim of the disk, the accretion spot sweeps across
first one face of the tilted disk and then the other during one orbit, 
  impacting the rim only twice during one cycle,
along the line of nodes.  Thus, the accretion mass and energy 
  is deposited throughout a range of radii in the disk, and
this may act to stabilize the disk against DN outbursts.  
The slow rise of the first 
dwarf nova oscillations after the long quiescent state 
 indicates that this is an inside-out outburst, consistent
with the deposition of mass primarily in the inner disk. 
  Clearly these data motivate further study of the
thermal-viscous limit cycle instability in tilted disks.

 The primary interpolation scheme we considered, given
 in eqn. (\ref{log}), has clear problems in making good
  superoutbursts. The model superoutbursts have shoulders
  which comprise about a third to half of the total viscous
  plateau, rather than the $\sim$10\% observed.
  However, for the H98 scaling with $T_0=12500$ K, the shoulders
  are appropriately short, $\sim$2 d, and the viscous plateaus much longer.
    The rate of decay of the viscous plateau
  is too fast, but that is because the input parameters have
  been optimized for the scaling given in eqn. (\ref{log}).

\section{Discussion}

In 
contrast to the thermal-tidal instability of Osaki (Ichikawa \& Osaki 1992, 1994;
Ichikawa, Hirose, \& Osaki 1993) in which the tidal torque is artificially
enhanced by a factor of 20 when the outer disk expands beyond the point
of 3:1 commensurability 
   with the orbital period,
   our
detailed modeling shows that the regular and superoutbursts can produced
with a constant set of parameters.   
  In  retrospect,
 one  may legitimately question
  the motivation for the increased
  tidal torque  accompanying the expansion of the disk,    
 considering  that the ultimate  driving  force behind  the tidal torque is the gravitational
field of the secondary, which remains  unchanged.
   We share a common viewpoint with van Paradijs (1983) in that
we regard the superoutbursts in the SU
UMa systems as an extension of the long outbursts in DNe
above the period gap, such as U Gem and SS Cyg (Cannizzo \& Mattei 1992, 1998;
C93b). 
    Continued monitoring of V344 Lyr 
    should reveal the stability of the long
term  sequencing, i.e., variations in the number of shorts sandwiched between
two longs.

 The shoulders on the rise to superoutburst in the models
 have more structure than the shoulder 
 in the observed superoutburst.
    The shoulders arise because the initial tendency of
the triggering of thermal instability  is to produce a
short outburst. However, due to the  long-term net build-up 
  of mass in the disk at large radii between successive short outbursts,
   there is actually enough material accumulated to support
  a superoutburst. It takes a little longer however for the
  outward moving  heating front to progress to  larger radii
   and to fully access this additional store of matter,
   which manifests itself as a ring of enhanced surface 
density in the outer disk.
    This  explains the fact that many observed superoutbursts
  appear to start as normal (short) outbursts.
 The subsequent rise after the shoulder is due to the 
 full radial extent of the disk out to the outer edge
 making a transition to the hot state.
 In contrast, for the short (normal) outbursts, 
 not enough disk mass has accumulated at larger radii
 by the time of the outburst
 to support the heating transition front
 propagating to large radii, therefore the
 heating front stalls at some $r < r_{\rm outer}$
 and is reflected as a cooling front,
 leading to pointy maxima rather than viscous plateaus.

Our testing has delineated a range
of allowable parameters:

(1) The logarithmic interpolation 
   for $\alpha$ produces
  well-separated outbursts, with concave downward decays,
  as observed.
  The  linear  interpolation produces
  outbursts with minimal quiescence and  concave upward decays.
   Given the physical underpinnings of the MRI, the logarithmic interpolation
  is better motivated since the MRI strength should depend
  on the partial ionization fraction, which is a strongly increasing
  function of temperature.

  In terms of the superoutburst,
   the H98 scaling with $T_0=12500$ K produces the best overall
   results. The shoulder feature only persists $\sim2$ d, as observed,
  and the viscous plateau is longer. About 70\% of the stored mass 
  is accreted. The main deficiencies are that the fast decay
(i.e., that following the viscous plateau) 
  is slower-than-exponential, and the rise times are too fast (as with 
the other calculations).

(2) The minimum number of grid  points necessary is $\ga10^3$
  (in accord with C93b and H98).
   Even for high $N$, the sequencing is not perfectly stable.

(3) Of the terms entering into the thermal energy equation, the 
  radial viscous transport term $C$ and the second advective term, 
  $v_r (\partial T/\partial r)$ appear to be the most important.
  The radial radiative flux term $D$ and the first advective term, 
  involving $\partial(rv_r)/\partial r$,  are of lesser importance.

(4) The values for $\alpha_{\rm cold}=0.0025$,  $\alpha_{\rm hot}=0.1$,
    and ${\dot M}_T=2\times 10^{-10}\msunyr$ are mandated
   by the observed spacing and duration of short outbursts
   and the superoutburst. Increasing their values decreases
   the outburst time scales.  
   The inferred values for  $\alpha_{\rm cold}$ and
    $\alpha_{\rm hot}$ are to be viewed merely
  as best-fit parameters for V344 Lyr; they do not allow us to draw any 
 deep conclusions as regards other systems.

(5) The values $r_{\rm inner}$ and $r_{\rm outer}$
   strongly mediate the number of short  outbursts
  between two superoutbursts.
  For $N=800$ and $f_t=1/160$
  we have $\sim$5 shorts between two longs
  for $r_{\rm inner}=2\times 10^9$ cm,
 or $\sim$3$R_{\rm WD}$ for $M_{\rm WD}=0.6\msun$.
    Observations by Howell et al. (1999) seem to favor 
  an evacuated inner disk. Several  possibilities are
   discussed by these workers, including the possibility
  of  weak intermediate polars in the SU  UMa systems.
  The outbursts in our models are triggered
  at intermediate disk radii,
  therefore the outer disk radius
   has minimal effect on the outburst recurrence times and durations.
  Increasing $r_{\rm outer}$ from $1\times 10^{10}$ cm to $2\times 10^{10}$ cm
  increases the number of shorts between to successive longs,
  and also distorts the  profile of the superoutburst.

With $\alpha_{\rm hot}=0.1$,  we match the rate of decay
  of the viscous portion of the superoutburst, and also find a
  $\sim$0.5 mag shoulder on the rise to superoutburst.
   It is not clear why previous detailed time dependent modelers
  did not find this shoulder (e.g., S04). 
The main deficiency in the models
  with the standard logarithmic 
  $\alpha_{\rm cold} \leftrightarrow \alpha_{\rm hot}$
   interpolation
is their failure to produce a sufficiently long  
   viscous plateau on the superoutburst.
  The observed plateau has a decay rate of $\sim$12 d mag$^{-1}$
  and lasts for $\sim$12 d, which corresponds to $\sim$1
$e-$folding in the decay of the disk mass, meaning about 60\% of
the disk mass should be accreted. 
  Our viscous plateau lasts $\sim$6$-$8 d, with proportionally less
mass accreted. 
   However, the light curve using the H98 scaling
  with $T_0=12500$ K  (which appears to be 
   physically motivated insofar as it tracks $\xi$) 
  has substantially
   longer superoutbursts, $\sim10$ d,
    with appropriately short shoulders as 
observed. The reason for the shorter shoulders
  is that the temperature within the outward moving 
  heating front accompanying superoutburst onset is  between about
  $10^4$ and $2\times 10^4$ K. 
  As can be seen in Figure 5 by comparing
the left-most green curve (i.e., the  H98\_12500 scaling)
    with the ``log'' curve, 
for the temperature range  $10^4$ K $ < T <  2\times 10^4$ K,
   $\alpha$  is larger in the  H98\_12500 scaling
and therefore the heating front travels faster since its 
speed varies with $\alpha$ (Menou, Hameury, \& Stehle 1999).
  The  duration of the shoulder feature corresponds to the interval
during which the heating front is moving outward and finding
that enough mass is present at large radii to support a superoutburst.
  This causes the dip and subsequent rise to the plateau of
the superoutburst.  

 The values for $\alpha_{\rm cold}$,  $\alpha_{\rm hot}$,
  and ${\dot M}_T$ have already been optimized to get the supercycle
and normal cycle periods of $\sim$100 d and $\sim$10 d, respectively.
  One suspects that the scalings for $\Sigma_{\rm max}(\alpha_{\rm cold})$ 
  and  $\Sigma_{\rm min}(\alpha_{\rm hot})$ 
    are too simplistic 
   and fail to take into account the actual disk physics,
   particularly that of the quiescent state, which is presently unknown. 
   One could envision introducing an ad hoc multiplicative parameter
  for  $\Sigma_{\rm max}$, for instance to artificially enhance it, 
   but then one would have to re-adjust all the other parameters, 
  in particular increasing  ${\dot M}_T$ which is already uncomfortably large.
   In view of these considerations we
  consider our experiment moderately successful in that we are able to
  reproduce the duration of the viscous plateau within a factor of two
   in an over-constrained system for which 
 $\alpha_{\rm cold}$ and  $\alpha_{\rm hot}$
   have already been determined by the outburst properties.

What about having enhanced mass transfer from the secondary
  prolong the duration of the superoutburst?
   Smak  (2004) presented observational evidence 
  for enhanced hot spot brightness 
  near superoutburst maxima
  which he interpreted as being due to 
 enhanced mass transfer brought about by the strong
irradiation of the secondary.
   Osaki \& Meyer (2003) presented a theoretical 
   study  of the effects of  irradiation
  on the photosphere of the secondary star in SU UMa systems
  and concluded that mass transfer could not be enhanced
  by a significant amount 
   because the Coriolis force prevents
   the formation of a circulation
   flow transporting heat toward L1.
   Viallet \& Hameury (2007$=$V07, 2008$=$V08)
    present two-dimensional, time dependent
  calculations of the surface flow of material of
irradiated secondaries  and find the physical situation to
   be significantly different from that envisioned by either 
    Osaki \& Meyer or Smak.  
  The Coriolis
force leads to the formation of a circulation flow
  from the high latitude regions to the vicinity of the L1 point,
  but rapid cooling of the gas as it enters the shielded equatorial
  region  prevents significant heat transport (V07).
  V07 conclude that any resulting mass transfer rate enhancement
is likely to be moderate and unable to account for the 
   duration of long outbursts in dwarf novae.
   V08 look   at two effects that could potentially
  induce secondary mass transfer rate variations in
   dwarf novae and soft X-ray transients: 
  irradiation of the secondary by the hot 
  outer disk rim during outburst,
  and scattered radiation by optically thin outflowing material.
   They conclude that for dwarf nova parameters the effects
  on enhanced mass transfer are negligible.
   For completeness we note that some of the assumptions
 of V07 and V08 may not apply if the disk is 
  warped or tilted,
  in which case the L1 point could in principle be directly irradiated.

%
  S04 presented
  detailed time dependent calculations of
  the accretion disk evolution
in  both the 
       TTIM  and the EMTM,
   and in the end concluded
  that the EMTM was favored. Their
arguments are rather general, however, and
they state ``... we have not proven the EMTM to be correct
   nor the TTIM not to work.''
     Both models have a number of free parameters 
in addition to those already present in the standard limit cycle model.
  For the TTIM one has the time delay between the attainment of 
the 3:1 radius by the outer disk edge and the onset of enhanced tidal
torque, in addition to the functional form and amplitude associated 
with the increase.
    Similarly the EMTM has as free parameters the degree of augmentation
of ${\dot M}_T$ during superoutburst, as well as its functional form
  with time.

In addition, there may be a problem with 
 the viscous decay in the EMTM:
   The V344 Lyr superoutburst maintained a closely exponential decay 
                                over $\sim$1 mag, whereas our models
   as well as others reveal a definite slower-than-exponential trend 
  (i.e., concave upward when plotted as magnitude versus time)
  in viscous plateaus  covering that much dynamic range. 
       Although the viscous decays can be lengthened in the EMTM, 
  they tend to have noticeable deviations from exponentiality, 
    in contrast to the V344 superoutburst.
  The fact that the short orbital period system WZ Sge did
show a        deviation from exponentiality
  in its 2001 outburst, which also spanned a greater  dynamic range
  than in the V344 Lyr superoutburst ($\sim2$ mag versus $\sim$1 mag),
  may indicate that irradiation-induced  enhanced mass transfer
is a factor at much shorter orbital periods than for V344 Lyr.
  In addition, the
  amplitude of the $\sim$0.5 mag shoulder on the rise to superoutburst
  is roughly matched in our calculations; 
   with enhanced mass transfer the shoulder amplitude 
  is too great ($\ga1$ mag).
   We consider it 
  fortunate that we  have
matched the decay as well as we have since the scalings for the
  extrema in the limit cycle model, 
    in particular $\Sigma_{\rm max}(\alpha_{\rm cold})$,
   are most probably 
         gross simplifications to a more complicated situation
     that is currently beyond our understanding.

\section{Conclusion}

We have presented detailed time
dependent calculations of the accretion disk
limit cycle mechanism with application to 
the {\it Kepler} light curve of V344 Lyr.
We find that the basic disk instability model
is able to account for the mixture of normal and superoutbursts
with a minimal amount of fine-tuning.
   Given the additional free parameters associated
   with  both the thermal-tidal model
   and the enhanced mass transfer models,
     we do not attempt to examine their possible ramifications
in this work. It may
be that the plain vanilla disk instability acting alone
  is sufficient to explain most aspects of 
the long term  V344 Lyr light curve.
 We also note that the normal
  outbursts in V344 Lyr do not show
a monotonic increase in recurrence time and  
   amplitude between successive superoutbursts
        as predicted by the TTIM (see
  Ichikawa, Hirose, \& Osaki 1993, Fig. 1;
                     Osaki 2005, Fig. 3),
  but rather a maximum in these quantities near the middle of 
a supercycle.    The models we present
  also do not show the observed variations.

 One can in some sense view the alternation
of many successive short outbursts with a superoutburst
as an extensive of the pattern of short and long
  outbursts seen in the longer orbital period
systems such as U Gem and SS Cyg (van Paradijs 1983);
  for SU UMa parameters one has many more  
  short outbursts between two longs
   than in a system like SS Cyg,
  which exhibits LSLSLS... most of the time 
   (Cannizzo \& Mattei 1992).
   For our best-fit model parameters,
 we find that $\ga 800$ grid points are
  required for numerical stability, 
  with a resulting pattern in which 5$-$7 short outbursts
 occur between two successive superoutbursts
  (for $f_t = 1/640$).
    Overall,
    our outbursts are at about the
   same brightness as those observed in V344 Lyr
   for an intermediately inclined disk.
   Having constrained $\alpha_{\rm cold}$ and
$\alpha_{\rm hot}$ so as to produce the overall
recurrence time scales of short outbursts and superoutbursts,
  there are several obvious deficiencies: the rise times
  of all outbursts are too fast, the superoutburst duration
 is too short, and the duration of the shoulder is too long
and has an extra scalloped-out feature prior to the 
observed shoulder. 
   However, the more physically motivated
   H98 scaling with $T_0=12500$ K does  a better job  on 
  making superoutbursts with minimal shoulders
  as observed, 
 lasting $\sim$2 d, 
 and also longer viscous plateaus,
    $\sim10$ d vs. the observed $\sim12$ d.
 The normal outburst embedded at the start of a superoutburst
is due to the initial triggering of the thermal instability and outward
 movement of the heating front; the subsequent dip and rise to superoutburst
maximum comes about because the heating front continues to move outward,
and subsequently to access the stored material at large radii in the disk
which accumulates where the heating front spike stalls at the end of
 a normal outburst.

  Even as we struggle to account for the finer points 
  of the light curve, such as the shoulder on the 
superoutburst, we feel compelled to step
 back briefly from our myopic scrutiny of the details
  and give some modicum of thanks and appreciation
  for the fact that the {\it Kepler} data
   represent such a huge advance over what was available
  even as recently as one  year ago. 
 For long term dwarf nova light curves, the only data sets
 that had been available were those  amassed by the AAVSO and similar
organizations. While these data have certainly been useful
 and yielded many interesting results, one could
 only study in detail the brightest systems, like SS Cyg and U Gem.
In addition, the data quality was uneven and  the time sampling coarse.
 When new data become available that are an
   order of magnitude better than the previous data,
  it becomes possible to make discoveries 
   that are different in ``kind'', not just in
 ``degree''.
   The {\it Kepler} data will provide a physical touchstone 
  for modelers with the potential to substantially
  improve our understanding of the physics of accretion disks;
  our work represents only the first small step in this regard.


\acknowledgements

We thank James Bubeck
  (under the SESDA II contract)
  at Goddard Space Flight Center 
 for assistance
  in providing local computational
  resources,
   and 
  Marcus Hohlmann and Patrick Ford from the 
  FIT High Energy Physics group and the Domestic Nuclear
Detection Office in the Dept. of Homeland Security for making 
 additional resources on a Linux cluster available.

We also thank the anonymous referee whose comments
   helped improve and clarify the paper.
We acknowledge the contributions
  of the entire {\it Kepler}  team.

\vfil\eject

\def\mnras{MNRAS}
\def\apj{ApJ}
\def\apjs{ApJS}
\def\apjl{ApJL}
\def\aj{AJ}
\def\araa{ARA\&A}
\def\aap{A\&A}
\def\aapl{A\&AL}
\def\pasj{PASJ}


\centerline{References}

\noindent
Ak, T., Bilir, S., Ak, S., \& Ezer, Z. 2008, New Astr, 13, 133

\noindent
Ak, T., Ozkan, M.~T., \& Mattei, J.~A. 2002, \aap, 389, 478

\noindent
Angelini, L., \& Verbunt, F. 1989, \mnras, 238, 697

\noindent
Balbus, S.~A., \& Hawley, J.~F. 1998, Rev Mod Phys, 70, 1

\noindent
   Barrett, P.,
        O'Donoghue, D., \& Warner, B.\ 1988, \mnras, 233, 759

\noindent
Bath, G.~T., \& Pringle, J.~E. 1981, \mnras, 194, 967

\noindent
Buat-M\'enard, V., \& Hameury, J.-M. 2002, \aap, 386, 891

\noindent
Buat-M\'enard, V., Hameury, J.-M., \& Lasota, J.-P. 2001, \aap, 366, 612

\noindent
Cannizzo, J.~K. 1993a, in Accretion Disks in
       Compact Stellar Systems,
   ed. J. C. Wheeler (Singapore: World Scientific), 6 (C93a)
%

\noindent
Cannizzo, J.~K. 1993b, \apj, 419, 318 (C93b)

\noindent
Cannizzo, J.~K. 1994, \apj, 435, 389

\noindent
Cannizzo, J.~K. 1998a, \apj, 493, 426

\noindent
Cannizzo, J.~K. 1998b, \apj, 494, 366

\noindent
Cannizzo, J.~K. 2001a, \apj, 556, 847

\noindent
Cannizzo, J.~K. 2001b, \apj, 561, L175

\noindent
Cannizzo, J.~K., Gehrels, N., \& Mattei, J.~A. 2002, \apj, 579, 760

\noindent
Cannizzo, J.~K., \& Mattei, J.~A. 1992, \apj, 401, 642

\noindent
Cannizzo, J.~K., \& Mattei, J.~A. 1998, \apj, 505, 344

\noindent
Cannizzo, J.~K., Shafter, A.~W., \& Wheeler, J.~C., 
    1988, \apj, 333, 227 (CSW)

\noindent
Cannizzo, J.~K., \& Wheeler, J.~C., 
    1984, \apjs, 55, 367

\noindent
Cannizzo, J.~K., Wheeler, J.~C., 
  \& Polidan, R.~S. 1986, \apj, 301, 634 (C86)

\noindent
Faulkner, J., Lin, D.~N.~C., \& Papaloizou, J.
     1983, \mnras, 205, 359

\noindent
   Foulkes, S.~B., Haswell,
        C.~A., \& Murray, J.~R.\ 2006, \mnras, 366, 1399

\noindent
Fromang, S., \& Papaloizou, J. 2007,
  \aap, 476, 1113

\noindent
Hameury, J.-M. 2002, in The Physics of Cataclysmic Variables
   and Related Objects, ed. B.~T. G\"ansicke, K. Beuermann, \& K. Reinsch
   (San Francisco; ASP), 377

\noindent
Hameury, J.-M., Menou, K., Dubus, G., Lasota, J.-P.,
    \& Hure, J.-M. 
      1998, \mnras, 298, 1048 (H98)

\noindent
Harrison, T.~E., Johnson, J.~J., McArthur, B.~E., Benedict, G.~F., Szkody, P.,
   Howell, S.~B., \& Gelino, D.~M. 2004, AJ, 127, 460

\noindent
Harrison, T.~E., McNamara, B.~J., Szkody, P., McArthur, B.~E.,
         Benedict, G.~F., Klemola, A.~R., \& Gilliland, R.~L.
     1999, \apjl, 515, L93

\noindent
Hirose, S., Krolik, J.~H., \& Blaes, O. 2009, \apj, 691, 16

\noindent
Hoffmeister, C., 1966, Astron. Nachr., 289, 139
%

\noindent
Howell, S.~B., Ciardi, D.~R., Szkody, P., van Paradijs, J., 
   Kuulkers, E., Cash, J., Sirk, M., \& Long, K.~S. 1999, PASP, 111, 342

\noindent
Howell, S.~B., Nelson, L.~A., \& Rappaport, S. 2001, \apj, 550, 897

\noindent
Ichikawa, S., Hirose, M., \&
    Osaki, Y. 1993, \pasj, 45, 243

\noindent
Ichikawa, S., \& Osaki, Y. 1992, \pasj, 44, 15

\noindent
Ichikawa, S., \& Osaki, Y. 1994, \pasj, 46, 621

\noindent
Kato, T. 1993, \pasj, 45, L67

\noindent
Kato, T., Poyner, G., \& Kinnunen, T. 2002, \mnras, 330, 53

\noindent
King, A.~R., Pringle, J.~E., \& Livio, M. 2007, \mnras, 376, 1740

\noindent
Kolb, U., King, A.~R., \& Ritter, H. 1998, \mnras, 298, L29

\noindent
Lasota, J.-P. 2001, New Astron. Rev., 45, 449

\noindent
Lightman, A.~P.
     1974, \apj, 194, 419

\noindent
Lin, D.~N.~C., Papaloizou, J., \& Faulkner, J.
     1985,  \mnras, 212, 105

\noindent
Lubow, S.~H. 1991a, \apj, 381, 259

\noindent
Lubow, S.~H. 1991b,  \apj, 381, 268

\noindent
Lubow, S.~H., \& Shu, F.~H. 1975, \apj, 198, 383

\noindent
Ludwig, K., \& Meyer, F. 1998, \aap, 329, 559

\noindent
Menou, K., Hameury, J.-M., \& Stehle, R. 1999, \mnras, 305, 79

\noindent
Meyer, F. \& Meyer-Hofmeister, E. 1981, \aap, 104, L10

\noindent
Meyer, F. \& Meyer-Hofmeister, E. 1982, \aap, 106, 34

\noindent
Meyer, F. \& Meyer-Hofmeister, E. 1984, \aap, 132, 143

\noindent
Meyer, F. \& Meyer-Hofmeister, E. 1994, \aap, 288, 175

\noindent
Mineshige, S.    1986, \pasj, 38, 831 (M86)

\noindent
Mineshige, S.    1987, Ap \& Sp. Sci., 130, 331 (M87)

\noindent
Mineshige, S. \& Osaki, Y. 1983, \pasj, 35, 377 (M83)

\noindent
Mineshige, S. \& Osaki, Y. 1985, \pasj, 37, 1 (M85)

\noindent
Mukai, K., Zietsman, E., \& Still, M. 
     2009, \apj, 707, 652

\noindent
Osaki, Y. 1989a, in Theory of Accretion Disks, ed. F. Meyer et al.
   (Kluwer: Dordrecht), 183

\noindent
Osaki, Y. 1989b, \pasj, 41, 1005

\noindent
Osaki, Y. 2005, Proc. Japan Acad., 81, 291

\noindent
Osaki, Y., \& Meyer, F. 2003, \aap, 401, 325

\noindent
Papaloizou, J., \& Pringle, J.~E. 1977, \mnras, 181, 441

\noindent
Pringle, J.~E. 1981, \araa, 19, 137

\noindent
Pringle, J.~E., Verbunt, F., \& Wade, R.~A.
   1986, \mnras, 221, 169 (P86)

\noindent
Schreiber, M.~R., Hameury, J.-M., \& Lasota, J.-P. 
        2003, A\&A, 410, 239 

\noindent
Schreiber, M.~R., Hameury, J.-M., \& Lasota, J.-P. 
        2004, \aap, 427, 621 (S04)

\noindent
Schreiber, M.~R., \& Lasota, J.-P. 
        2007, \aap, 473, 897

\noindent
Shakura, N.~I., \& Sunyaev, R.~A. 1973, \aap, 24, 337

\noindent
Simpson, J.~C., \& Wood, M.~A. 1998, \apj, 506, 360

\noindent
Sion, E.~M. 1995, \apj, 438, 876

\noindent
Smak, J. 1984, Acta Astr., 34, 161 (S84)

\noindent
Smak, J. 1993, Acta Astr., 43, 101

\noindent
Smak, J. 2004, Acta Astr., 54, 221

\noindent
Smak, J. 2009a, Acta Astr., 59, 89

\noindent
Smak, J. 2009b, Acta Astr., 59, 103

\noindent
Smak, J. 2009c, Acta Astr., 59, 109

\noindent
Smak, J. 2009d, Acta Astr., 59, 121

\noindent
Smak, J. 2010, Acta Astr., 60, 83

\noindent
Still, M.~R., Howell, S.~B., Wood, M.~A., Cannizzo, J.~K., \& Smale, A.~P.
        2010, \apjl, 717, 113

\noindent
van Paradijs, J. 1983, \aapl, 125, L16


\noindent
Viallet, M., \& Hameury, J.-M. 2007, \aap, 475, 597

\noindent
Viallet, M., \& Hameury, J.-M. 2008, \aap, 489, 699

\noindent
Vishniac, E.~T. 2009, \apj, 696, 1021 

\noindent
Warner, B. 
  1995a, Astrophys. \& Sp. Sci., 226, 187 

\noindent
Warner, B. 
  1995b, Cataclysmic Variable Stars (Cambridge: Cambridge Univ. Press)

\noindent
Warner, B., \& O'Donoghue, D. 1988, \mnras, 233, 705 

\noindent
Wheatley, P.~J., Mauche, C.~W., \& Mattei, J.~A. 
        2003, MNRAS, 345, 49

\noindent
Whitehurst, R. 1988, \mnras, 232, 35
 
\noindent
        Wood, M.~A., \& Burke, C.~J.\
        2007, \apj, 661, 1042

\noindent
      Wood, M.~A., Montgomery, M.~M., \&
        Simpson, J.~C.\ 2000, \apjl, 535, L39

\noindent
      Wood, M.~A., Thomas, D.~M., \&
        Simpson, J.~C.\ 2009, \mnras, 398, 2110


\clearpage

\pagestyle{empty}
\setlength{\voffset}{20mm}

\centering
\begin{table}
\centering
\vskip -6 truecm
\centerline{Table 1: Duration and Amplitude of Outbursts}
\tablewidth{0pc}
\tablenum{1}
\begin{tabular}{c c c}
\hline\hline
 Outburst & Duration (day) & Amplitude (mag) \\
\hline
1 & 4.9 & 4.0 \\
2 & 3.9 & 3.5 \\
3 & 4.4 & 3.6 \\
4 & 16.7 (SO)  & 4.5 \\
5 & 2.8  & 2.5\\
6 & 3.0  & 2.6\\
7 & 2.6  & 3.0\\
8 & 3.6  & 3.5\\
9 & 3.9  & 3.8\\
10 & 3.8  & 3.7\\
11 & 4.1  & 3.7\\
12 &17.1 (SO)  & 4.4\\
13 & 3.8  & 2.7\\
14 & 3.5  & 3.3\\
15 & 3.6  & 3.5\\
16 & 5.5  & 3.6\\
17 & 4.1  & 3.8\\
18 & 4.1  & 3.5\\
19 & 4.3  & 3.5\\
\hline
\end{tabular}
\end{table}

\clearpage


\vskip -0.75truein
\voffset=-0.25truein
\centerline{\bf Figure Captions}

\figcaption{
A comparison of the Johnson $V$ band filter ({\it red})
   with the {\it Kepler}
  filter ({\it blue}).
}

\figcaption{
{\it Kepler} light curve of V344 Lyr
     over 270 d showing outbursts
    every $\sim$10 d, with two superoutbursts.
  The flux is measured in $e^{-}$ cadence$^{-1}$,
where the cadence (time between integrations)
   is 1 m. The full time resolution is shown.
  To obtain visual magnitude, a correspondence 
   of $Kp=12$ to $10^7$ $e^{-}$ cadence$^{-1}$
  was adopted. (This conversion assumes an unreddened G2 star.)
}

\figcaption{
  The effect of using a
  logarithmic interpolation (eqn. [\ref{log}])
  versus  a linear one (eqn. [\ref{lin}])
    for $\alpha$.
    For these runs $N=800$ and $f_t=1/160$.
  Shown are the light curves
 for the logarithmic case 
   ({\it top panel})
  and 
   the linear case
   ({\it second panel}), 
   the accompanying 
  disk masses ({\it third and fourth panels}),
  and the radius of the hot/cold interface 
     $r_{\rm HC,10}$ during times
  when a transition front is present
          ({\it fifth and sixth panels}).
  For comparison the light curves for the disk-only
  contribution (i.e., excluding secondary and hot spot)
  are   shown as dotted curves in the first and second panels.
  The disk masses are in units of $10^{22}$ g
  and $r_{\rm HC,10}$ is in units of $10^{10}$ cm. 
}

\figcaption{
  The effect of using 
  the H98
  logarithmic factor $f$  (eqn. [\ref{hameury}]).
    For these runs $N=800$ and $f_t=1/160$.
  Shown are the light curves
 for $T_0=25000$ K
   ({\it top panel}),
     $18750$ K
   ({\it second panel}),
       and
     $12500$ K
   ({\it third panel}),
  the accompanying 
  disk masses ({\it fourth through sixth panels}),
   and the radius of the hot/cold interface
     $r_{\rm HC,10}$ during times
  when a transition front is present
          ({\it seventh through ninth panels}).
   The disk masses are in units of $10^{22}$ g
  and $r_{\rm HC,10}$ is in units of $10^{10}$ cm.
}

\figcaption{
  Shown are the different 
   $\alpha_{\rm cold}-\alpha_{\rm hot}$
   interpolations. 
   The blue lines show the scalings for 
   $T(\Sigma_{\rm max})$ 
   and 
   $T(\Sigma_{\rm min})$.
   The black curves indicate the 
 linear (eqn. [\ref{lin}]) 
  and
   logarithmic (eqn. [\ref{log}]) 
    interpolations.
   The green curves show the H98 scaling
   (eqn. [\ref{hameury}]), where the 
  three curves (left to right)
  represent  
    $T_0=25000$ K,
    $18750$ K,
     and
    $12500$ K.
   The three red curves show a measure of 
   the partial ionization; they are the relative contribution
   of electron pressure to total pressure as shown in Fig. 16 of C93b,
   for $\rho$(g cm$^{-3}$)$= 10^{-8}$, $10^{-7}$, and $10^{-6}$ (left to right).
   The values given in C93b have been multiplied by $2\alpha_{\rm hot}$ to take into account 
the fact that (1)  $P_e/P\rightarrow 0.5$ in the limit of complete ionization, 
    and (2) in the high temperature
   limit  $\alpha\rightarrow \alpha_{\rm hot} = 0.1$.
}
 


\figcaption{
Model light curves
  showing the detail of a superoutburst
  for separate calculations in which
   $N=   
       200$  ({\it short dashed}),
       400  ({\it  long dashed}),
       800  ({\it short dash-dotted}),
 and 1600  ({\it long  dash-dotted}).
   The green curve shows a superoutburst from the 
$T_0=12500$ K H98 scaling light curve of Fig. 4.
 For comparison the averaged {\it Kepler} data
     for the superoutbursts in Fig. 2 is shown ({\it first: blue},
                                                {\it second: magenta}),
   where the original $\Delta t=1$ min cadence
has been block-averaged  to $\Delta t=4$ hr and
a distance of 620 pc was assumed in converting from $Kp$ to $M_{\rm Kp}$.
  The two slanted red lines indicate a decay slope
  of 12 d mag$^{-1}$.
 The second panel shows the disk mass (in $10^{22}$ g).  
}

\figcaption{
  The effect of turning off different
   terms in the energy equation. 
 For these runs $N=800$ and $f_t=1/160$.
  Shown are light curves 
using: the full equation ({\it top panel}),
    the turbulent transport $C=0$ ({\it second panel}),
    the radiative transport $D=0$ ({\it third panel}),
   both  $C=0$ and  $D=0$ ({\it fourth panel}),
  the first advective term set to zero ({\it fifth panel}),
   and
  the second advective term set to zero ({\it sixth panel}).
}


\figcaption{
  The effect of the inner disk radius.
 For these runs $N=800$ and $f_t=1/160$.
  The four panels show the
   resultant light curves for
$r_{\rm inner}  = 2\times 10^9$ cm ({\it top panel}),
               $1.5\times 10^9$ cm ({\it second panel}),
                 $1\times 10^9$ cm ({\it third panel}),
        and   $ 0.5\times 10^9$ cm ({\it fourth panel}).
}

\figcaption{
  The effect of the outer disk radius.
 For these runs $N=800$ and $f_t=1/160$.
  The six  panels show the
   resultant light curves for 
$r_{\rm inner}  = 0.8\times 10^{10}$ cm    ({\it top panel}),
                 $1.0\times 10^{10}$ cm ({\it second panel}),
                $1.25\times 10^{10}$ cm  ({\it third panel}),
                $ 1.5\times 10^{10}$ cm ({\it fourth panel}),
               $ 1.75\times 10^{10}$ cm ({\it  fifth panel}),
          and   $ 2.0\times 10^{10}$ cm ({\it  sixth panel}).
}

\clearpage
\begin{figure}
\hskip -8 cm
\vskip -2 cm
\plotone{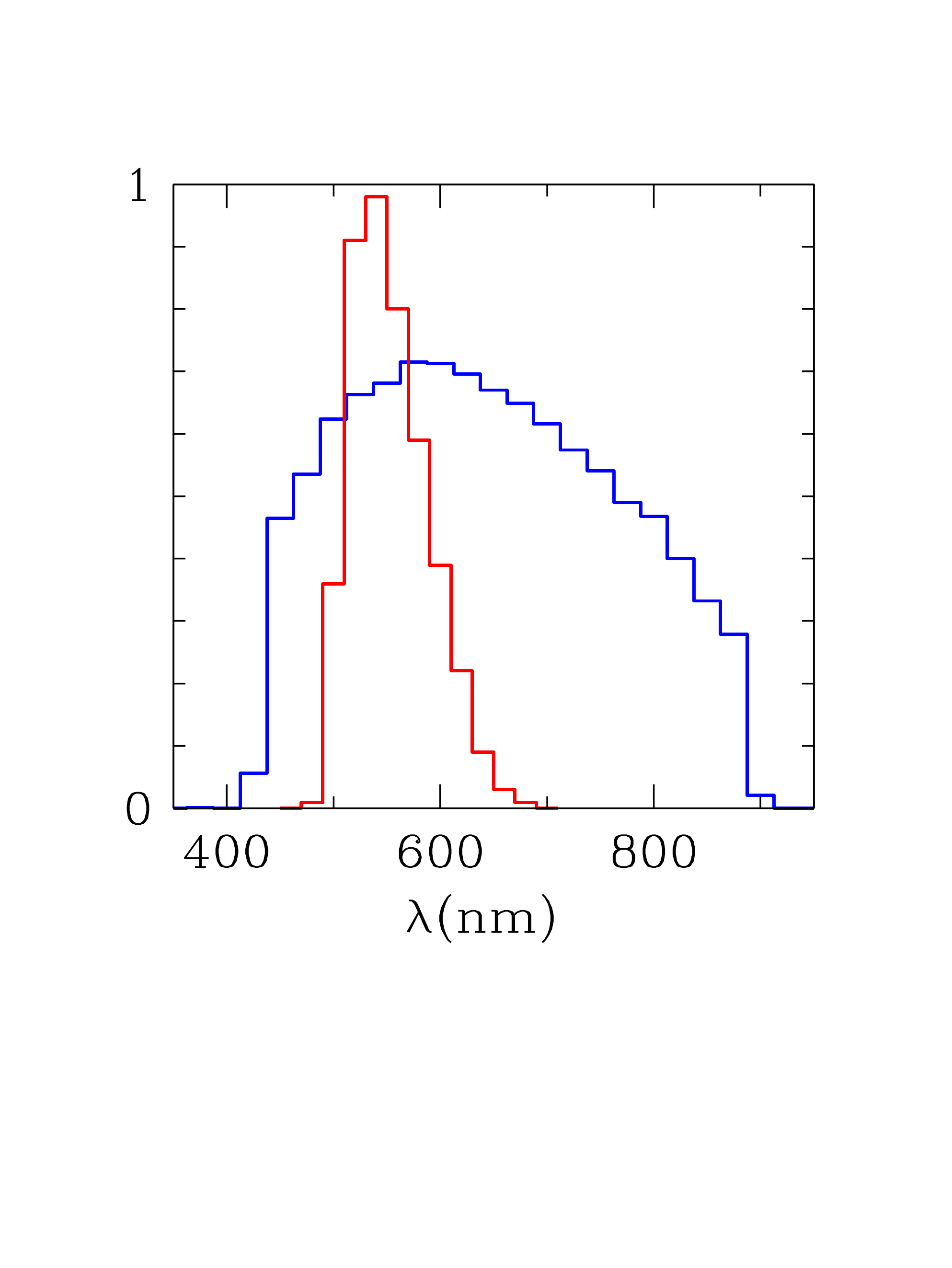}
\vskip -3 cm
\centerline{f1.ps}
\label{figV}
\end{figure}
\clearpage

\begin{figure}
\hskip -3.5 cm
\epsscale{.725}
\plotone{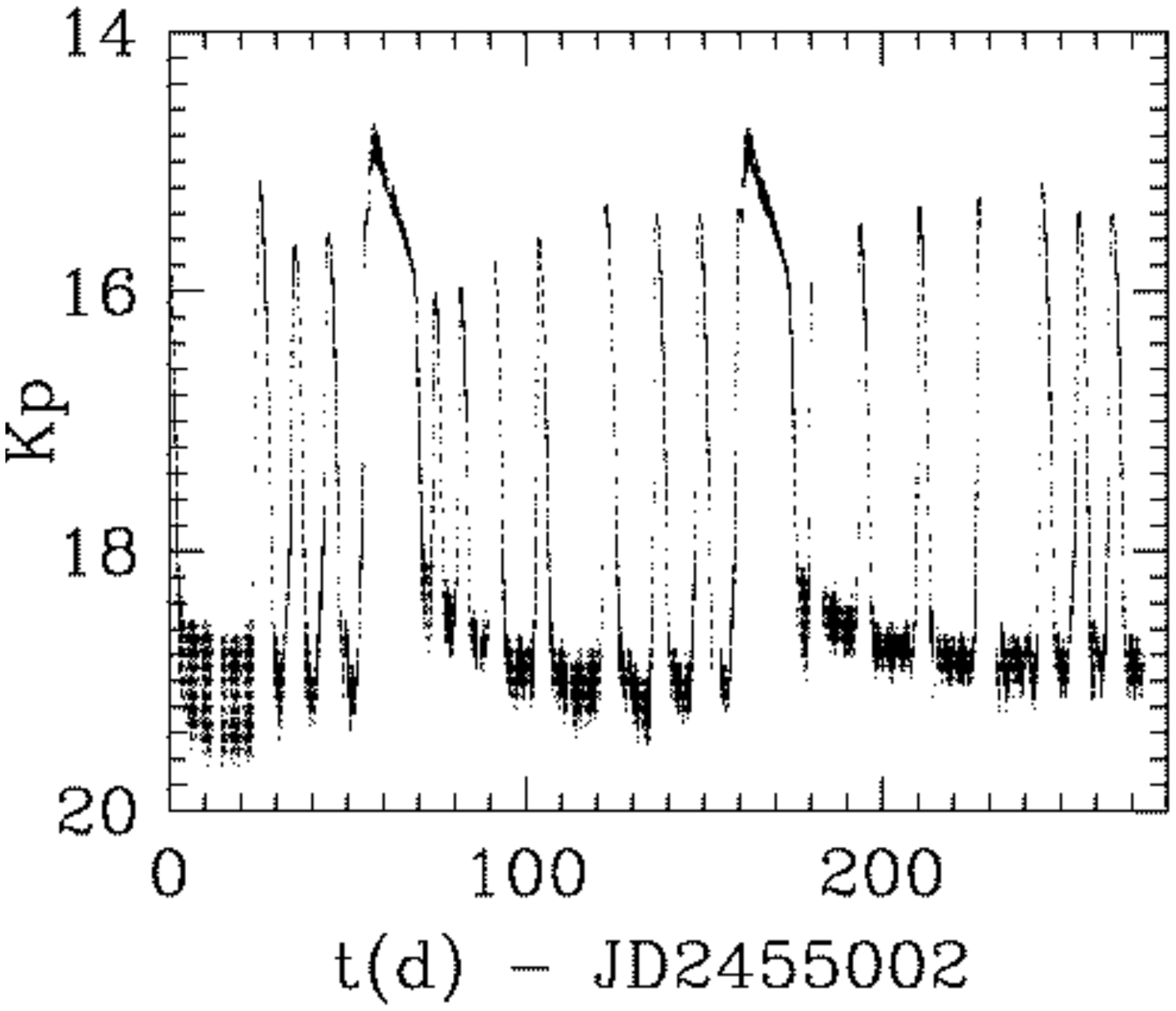}
\vskip  -5 cm
\centerline{f2.ps}
\label{fig_log_lin}
\end{figure}
\clearpage

\begin{figure}
\hskip -3.5 cm
\plotone{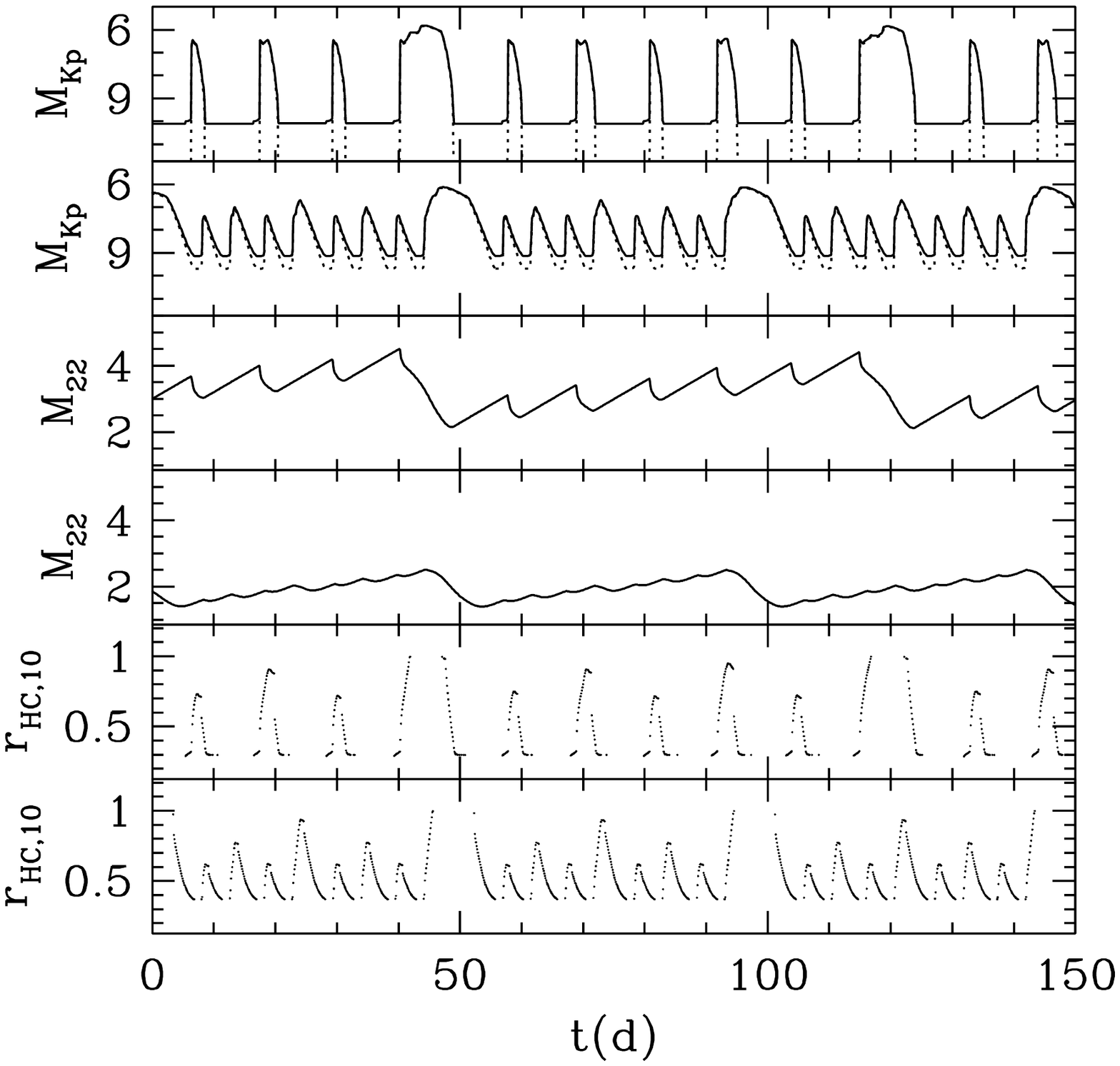}
\vskip  -6.5 cm
\centerline{f3.ps}
\label{fig_hameury}
\end{figure}
\clearpage

\clearpage
\begin{figure}
\hskip -3.5 cm
\plotone{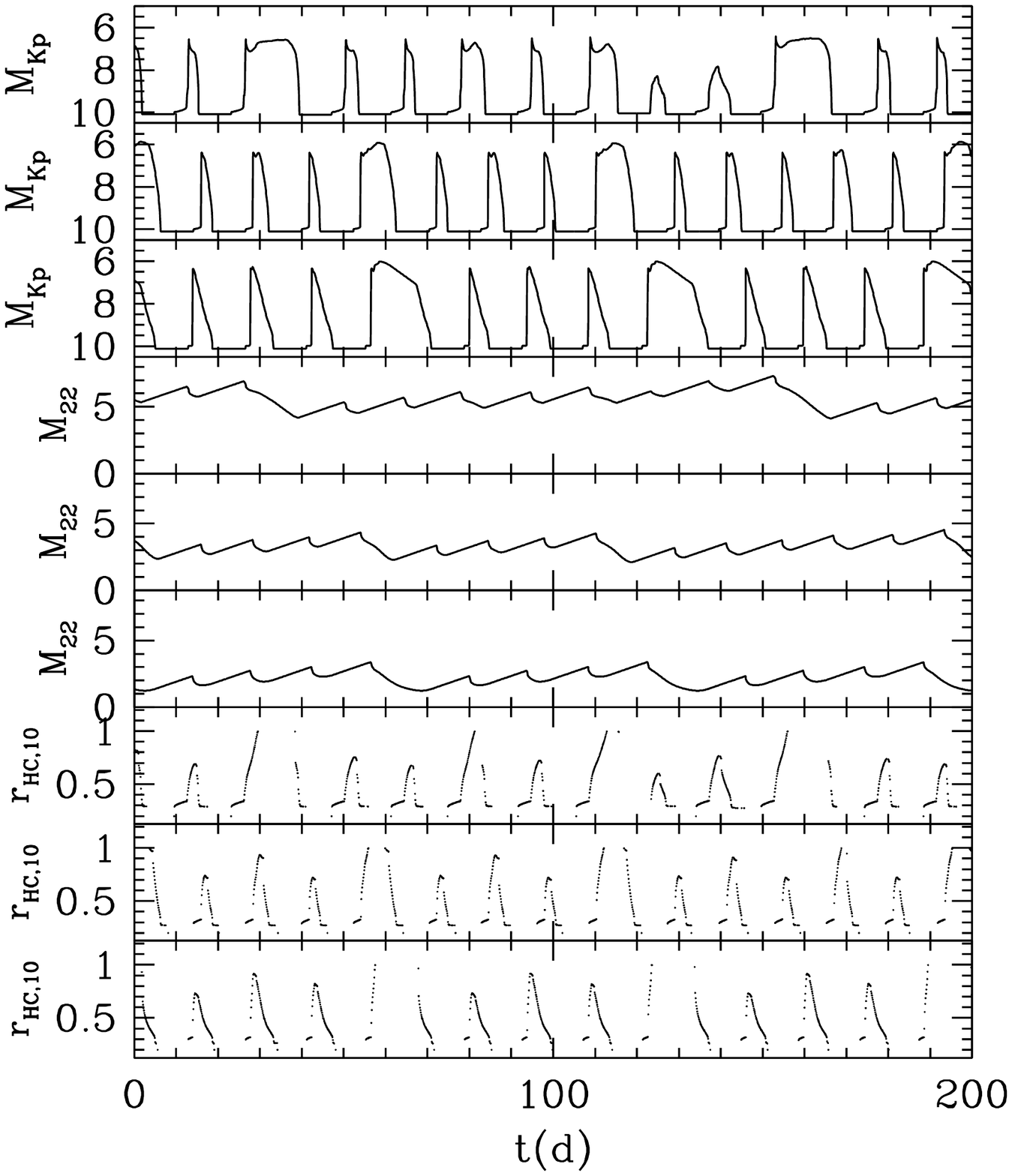}
\vskip  -4.5 cm
\centerline{f4.ps}
\label{fig_alp}
\end{figure}
\clearpage

\clearpage
\begin{figure}
\hskip -3.5 cm
\plotone{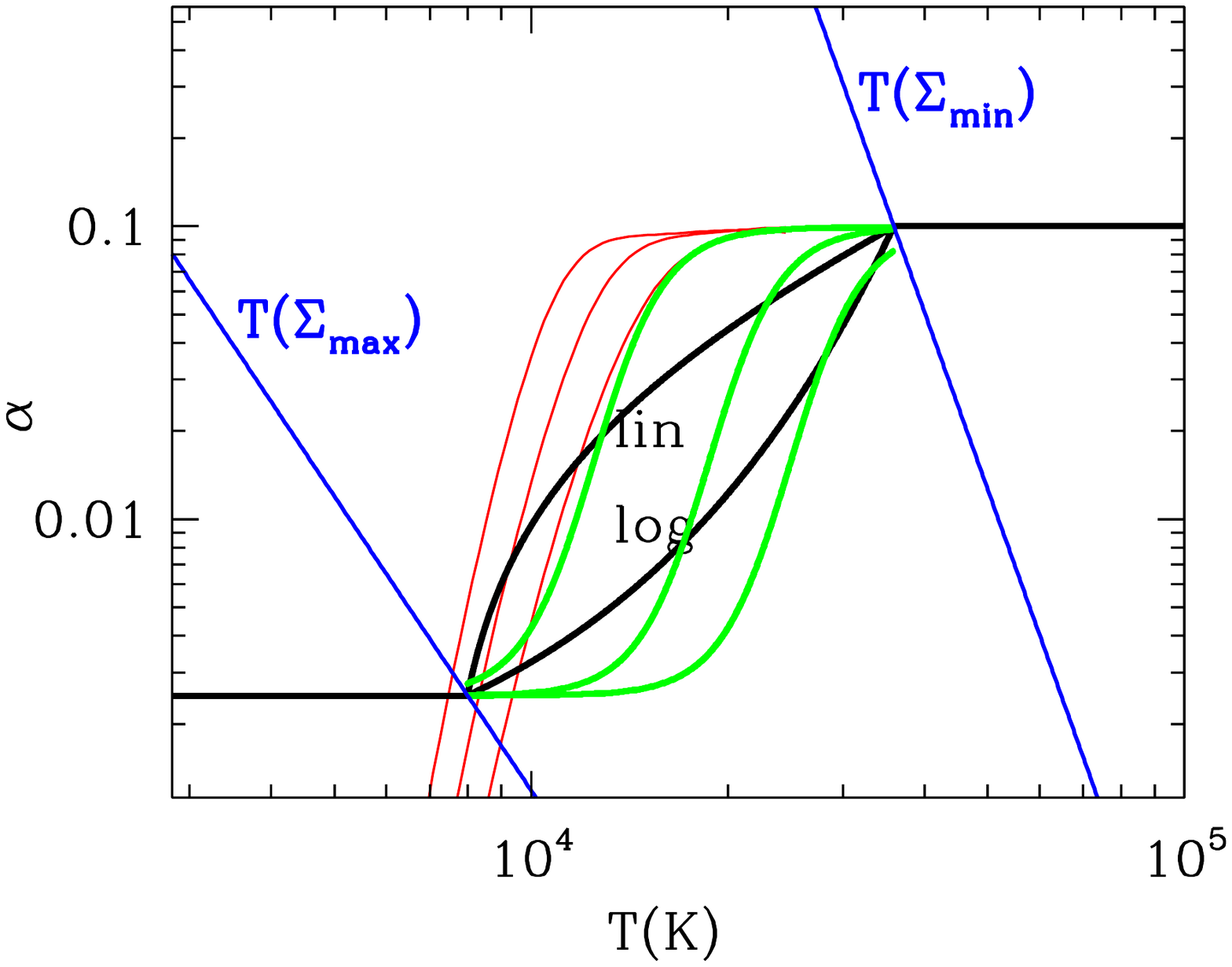}
\vskip  -6.5 cm
\centerline{f5.ps}
\label{fig_detail}
\end{figure}
\clearpage

\begin{figure}
\hskip -3.5 cm
\plotone{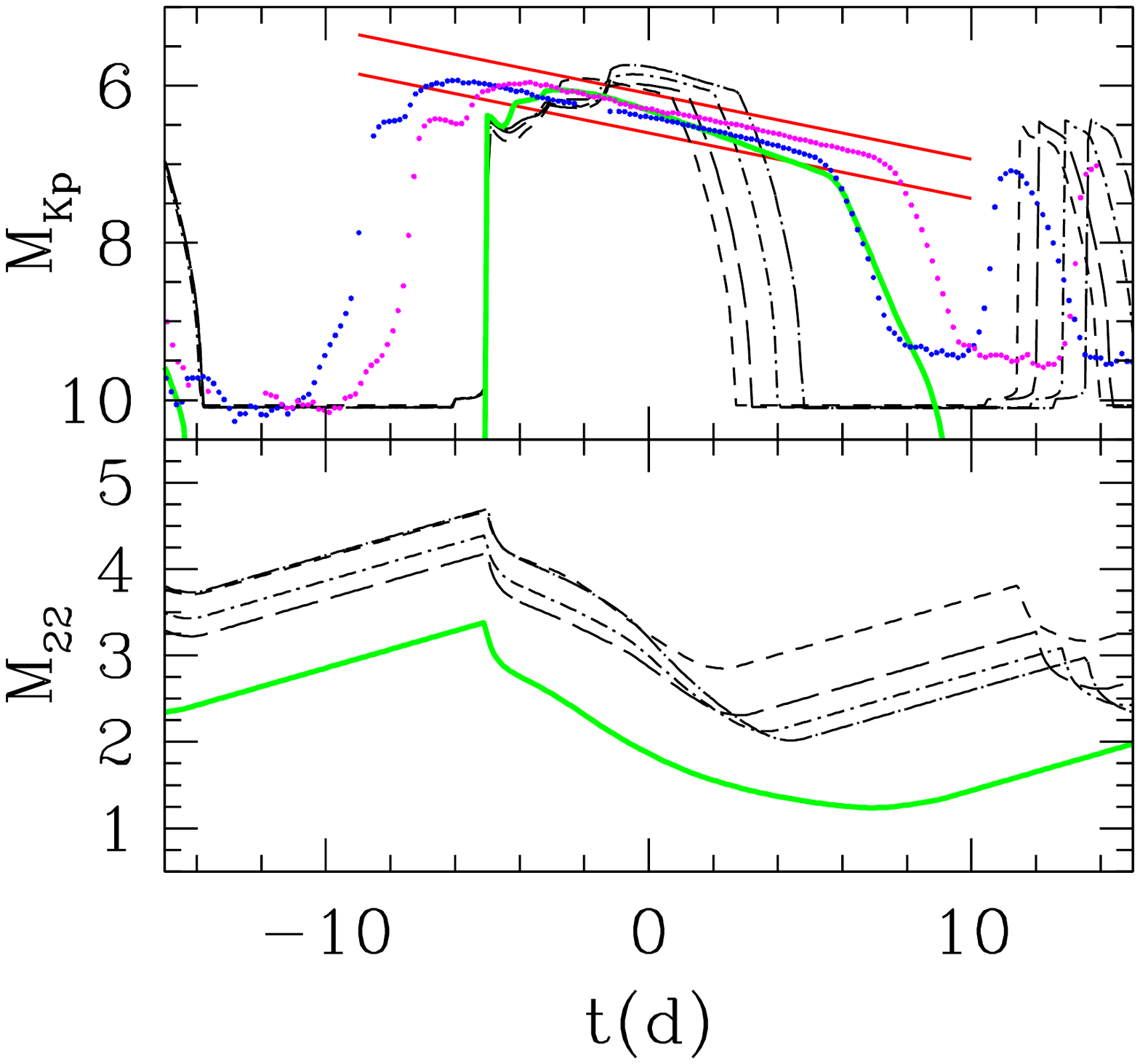}
\vskip  -6.5 cm
\centerline{f6.ps}
\label{fig_e}
\end{figure}
\clearpage

\clearpage
\begin{figure}
\hskip -3.5 cm
\plotone{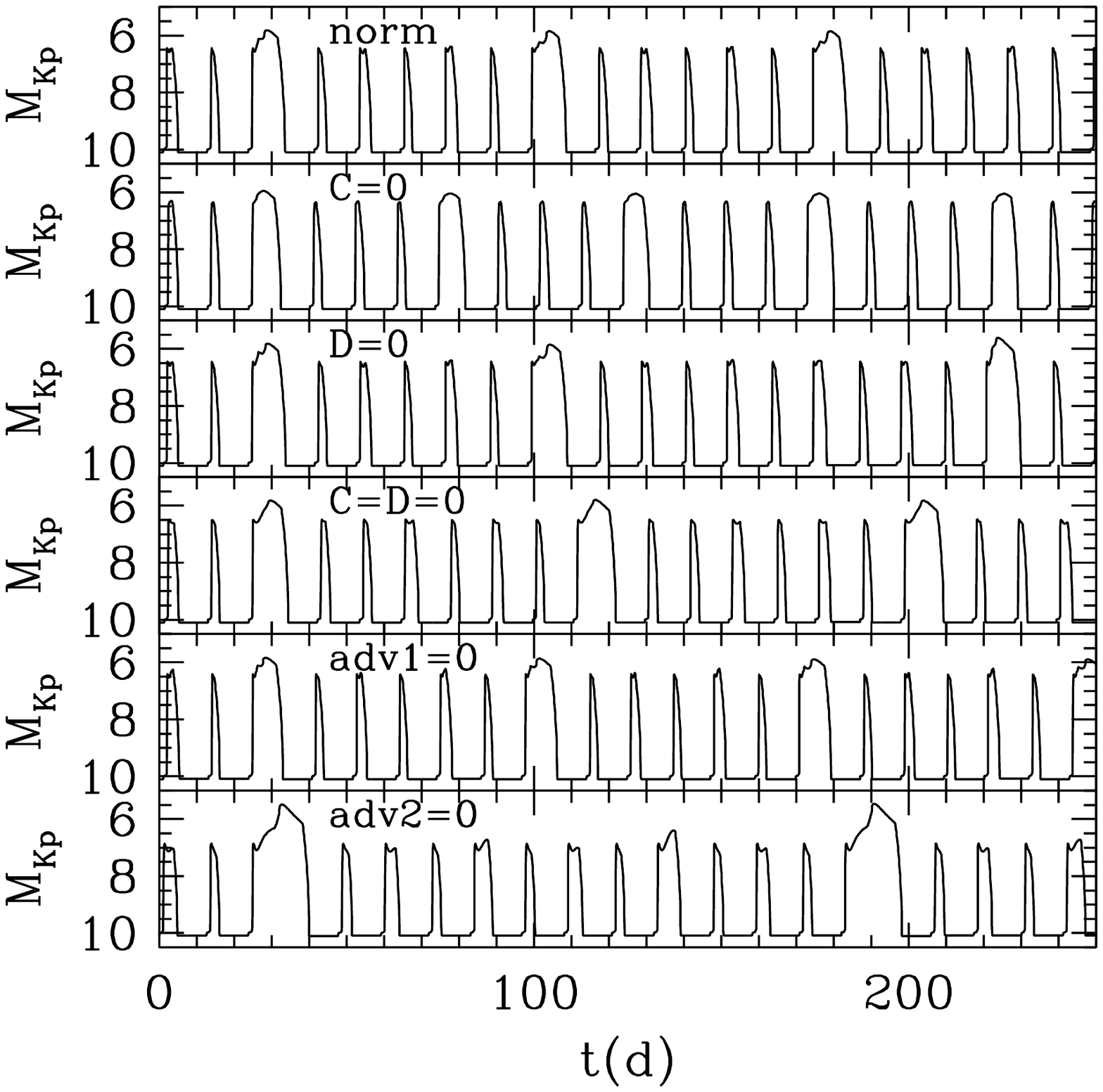}
\vskip  -6.5 cm
\centerline{f7.ps}
\label{fig_r1}
\end{figure}
\clearpage

\begin{figure}
\hskip -3.5 cm
\plotone{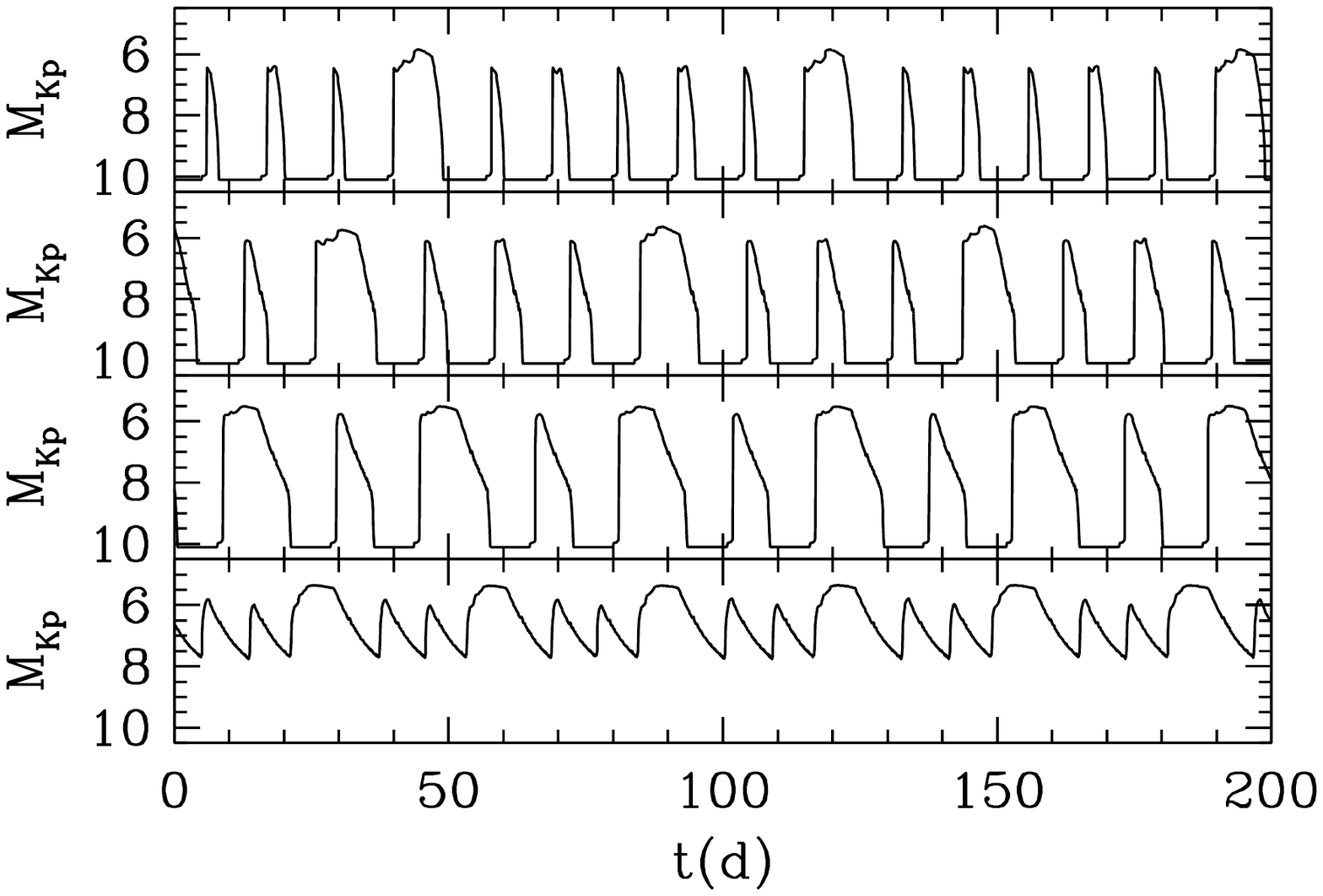}
\vskip  -6.5 cm
\centerline{f8.ps}
\label{fig_r2}
\end{figure}
\clearpage

\begin{figure}
\centering
\vskip -4.5 cm
\includegraphics{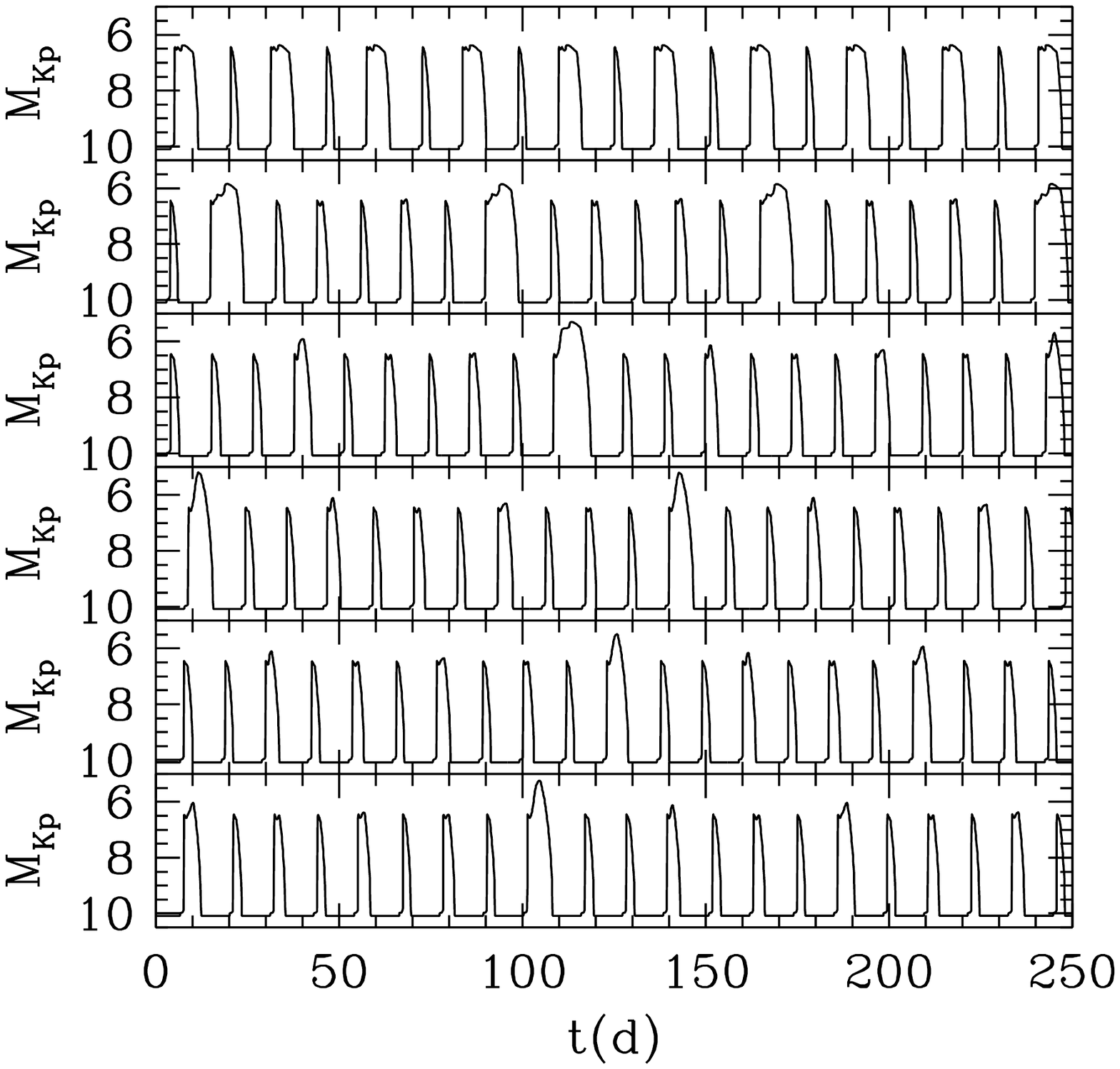}
\vskip -6.5 cm
\centerline{f9.ps}
\label{fig_r2}
\end{figure}
\clearpage

\end{document}